\documentclass[12pt]{article}
\usepackage[normalem]{ulem}
\usepackage{jstyle}
\usepackage{amsthm,mathrsfs,stmaryrd,bm}
\usepackage[vcentermath]{youngtab}
\usepackage{simplewick}
\usepackage{framed,pifont}
\usepackage{cancel}
\usepackage{verbatim}

\usepackage{diagbox}

\usepackage{tensor}

\definecolor{myred}{rgb}{1.0, 0.45, 0.400}

\usepackage{graphics}
\usepackage{graphicx}
\usepackage{epic}
\usepackage{eepic}

\def\g{\gamma}

\def\d{\delta}

\def\e{\epsilon}

\def\m{\mu}
\def\n{\nu}

\usepackage{latexsym}
\usepackage{amsmath}
\usepackage{wasysym}
\usepackage{amsfonts}
\usepackage{pifont}
\usepackage{bm}
\usepackage{epsf}
\usepackage{epsfig,graphicx}
\usepackage{amsmath,amssymb,bm,epsf,epsfig,graphicx}
\usepackage{diagbox}

\usepackage{color}
\definecolor{rougef}{rgb}{0.56,0,0}
\definecolor{vertf}{rgb}{0,0.5,0}
\definecolor{bleuf}{rgb}{0,0,0.8}
\definecolor{violetf}{rgb}{0.5,0,0.5}

\usepackage[vcentermath]{youngtab}

 \csname
@addtoreset\endcsname{equation}{section}

\def\3s{{s \choose 3}}
\def\4s{{s \choose 4}}
\def\5s{{s \choose 5}}
\def\6s{{s \choose 6}}

\def\12{\dfrac{1}{2}}

\def\ft{\footnote}

\def\2{\ell_2}

\def\be{\begin{equation}}
\def\ee{\end{equation}}
\def\bea{\begin{eqnarray}}
\def\eea{\end{eqnarray}}
\def\ba{\begin{array}}
\def\ea{\end{array}}
\def\bec{\begin{center}}
\def\ec{\end{center}}

\def\mth{\mathcal}

\def\beal{\begin{equation}\begin{aligned}}
\def\eeal{\end{aligned}\end{equation}}

\def\g{\gamma} 

\def\d{\delta} 

\def\e{\epsilon}

\def\m{\mu}
\def\n{\nu}

\def\cA{{\cal A}}

\def\cD{{\cal D}}

\def\cF{{\cal F}}

\def\cO{{\cal O}}

\author[a,1]{Andrea Campoleoni,}
\note{Research Associate of the Fund for Scientific Research -- FNRS, Belgium.}
\author[b, c]{Dario Francia}
\author[d]{and Carlo Heissenberg}

\affiliation[a]{Service de Physique de l'Univers, Champs et Gravitation\\
Universit{\'e} de Mons\\
Place du Parc, 20
B-7000 Mons, Belgium} 
\affiliation[b]{Centro Studi e Ricerche E. Fermi\\
Piazza del Viminale, 1 I-00184 Roma, Italy}
\affiliation[c]{Roma Tre University and INFN \\ Via della Vasca Navale, 84 I-00146 Roma, Italy}
\affiliation[d]{Scuola Normale Superiore and INFN,\\ Piazza dei Cavalieri, 7 I-56126 Pisa, Italy} 
\emailAdd{andrea.campoleoni@umons.ac.be}  \emailAdd{dario.francia@roma3.infn.it}
\emailAdd{carlo.heissenberg@sns.it}

\title{
\LARGE{On electromagnetic and colour memory\\ in even dimensions}}

\abstract{We explore memory effects associated to both Abelian and non-Abelian radiation getting to null infinity, in arbitrary even spacetime dimensions. Together with classical memories, linear and non-linear, amounting to permanent kicks in the velocity of the probes, we also discuss the higher-dimensional counterparts of quantum memory effects, manifesting themselves in modifications of the relative phases describing a configuration of several probes. In addition, we analyse the structure of the asymptotic symmetries of Maxwell's theory in any dimension, both even and odd, in the Lorenz gauge.}


\begin{document}

\maketitle

\setcounter{tocdepth}{2}

\tableofcontents
\newpage

%%%%%
\section{Introduction and Outlook} \label{sec: intro}
%%%%%
Memory effects are permanent changes in physical quantities pertaining to probes facing the passage of a burst of radiation close to or at null infinity. In this paper we illustrate various types of memories concerning matter probes in interaction with electromagnetic or Yang-Mills radiation, in arbitrary even-dimensional spacetimes.

 Following earlier works concerning gravitational memories \cite{ZP, Christodoulou:1991cr}, Bieri and Garfinkle \cite{BG2013} first pointed out the existence of analogous observables for electrically  charged matter in $D=4$. For charged test particles close to null infinity, they identified the relevant observable in the form of a ``velocity kick'' that the probes would experience as a consequence of the passage of a radiative perturbation. They also observed, in the same context, that the possibility to source those effects either with matter that does not reach null infinity or with massless sources allows one to identify two types of effects, termed ``linear'' (or ``ordinary'') in the former case and ``nonlinear'' (or ``null'') in the latter.

   Memory effects caused by electromagnetic fields were further discussed in  \cite{Tolish:2014bka} and later connected to asymptotic symmetries of four-dimensional Maxwell's theory, and thus identified as the classical counterparts of soft theorems, in \cite{Pasterski:2015zua, MaoetAl}. See also \cite{Sarkkinentesi, Sarkkinenpaper}. Moreover, the possible existence of analogous effects in dimension four or higher was investigated from various perspectives, with different (and sometimes not fully compatible) outcomes in \cite{Garfinkle:2017fre, Mao-Ouyang, HamadaSeoShiu, WaldOddD, HamadaSugishita, ShahinString, SatishchandranWald}. Those works adopted various approaches, focussed either on the classical solutions to the corresponding wave equations  or on the perturbative expansion of the gauge potentials with suitably assigned boundary  behaviour.   
   
  One of the main goals of this work is to describe ordinary and null memory effects for both Maxwell and Yang-Mills theories in even-dimensional spacetimes. To this end, we inspect the perturbative behaviour of the corresponding potentials in the Lorenz gauge and connect it to underlying residual gauge symmetries.

Following \cite{Garfinkle:2017fre, Mao-Ouyang}, we begin in Section \ref{sec:memories} by analysing the classical fields generated by specific background currents, so as to construct a number of concrete examples with which to compare our general results of the ensuing sections. In particular, we compute the null memory induced by a charged particle moving at the speed of light on a test particle initially at rest in even dimensions. In Section \ref{sec:em} we discuss various types of memories for Maxwell's theory in even-dimensional spacetimes. Upon employing a recursive gauge-fixing procedure, analogous to the one adopted in \cite{Strominger2} in the gravitational setup, we clarify the interpretation of both ordinary and null memory in any even $D$ as a residual gauge transformation acting at order $r^{4-D}$. We generalise this procedure to the non-Abelian case in Section \ref{sec:color}, where we obtain similar results and interpretation for linear and non-linear memory effects concerning matter in interaction with Yang-Mills radiation.

Both ordinary and null memory effects concern the permanent modifications affecting kinematical properties of the probes (the velocity in the case of spin-one radiation, the relative displacement in the gravitational case) due to the interaction of the latter with a transient radiation field close to null infinity. On the other hand, the interaction itself takes place in force of the matter probes possessing the required quantum numbers, electric charge or colour charge, whose overall configuration, whenever several particles are involved, can only be defined by providing the value of the corresponding connection field on the celestial sphere. In this sense, one should expect that the configuration itself be sensitive to the passage of radiation whenever the latter entails a vacuum transition between connections that differ by a (large) gauge transformation, even in the idealised  case of unperturbed kinematical variables ({\it i.e.}\ charges kept fixed at specified angles on the sphere). We refer to these types of memory effects as ``phase memory'' for the electromagnetic case and ``colour memory'' in the non-Abelian setting, to distinguish them from the memory effects affecting kinematical properties of the probes, that we collectively term simply as ``electromagnetic'' and ``Yang-Mills'' memories, respectively. To our knowledge, in $D=4$ phase memory was first discussed in  \cite{Susskind:2015hpa} while colour memory was analysed in  \cite{StromingerC}. Here we explore these phenomena in the higher, even-dimensional case, in Sections \ref{ssec:phasemem} and \ref{ssec:colourmem}, respectively. Let us observe that, differently from the ordinary and null memories that are purely classical effects, phase memory and colour memory naturally arise at the quantum level. It would be interesting to explore possible gravitational counterparts of the colour memory effect.

Memory effects in $D > 4$ appear at a subleading order with respect to radiation. Correspondingly, their counterparts at the level of gauge transformations do not comprise properly-defined infinite-dimensional asymptotic symmetries. The very existence of the latter, on the other hand, was sometimes doubted in the literature because of the need for too-slow falloffs on the gauge potentials that could in principle lead to divergences in physically sensible quantities. The asymptotic structure of spin-one gauge theories, in particular in connection to soft theorems was widely investigated in the literature \cite{Strominger_YM, BarnichYM, ACD2, Strominger_QED, StromingerQEDevenD, Campiglia, StromingerMagnetic, Avery_Schwab, soft_QED_Strominger, Henneaux:2018gfi, Hirai-Sugishita, ShahinMagnetic, Campoleoni:2018uib,Tristan}. More recently, the issue concerning the higher-dimensional extensions of those results was  further explored in a number of works \cite{MitraHePhoton, MitraHeGauge, MarcCedric-anyD, He:2019esa}. On our side, in Section \ref{sec:poly} we provide an analysis of the asymptotic symmetries of Maxwell's theory in any $D$, even and odd, stressing analogies and differences between the two cases, always working in the Lorenz gauge. In addition, we describe an explicit procedure for defining the surface charges on $\mathscr I$. 

In the appendices we collect a summary of our conventions (Appendix \ref{app:conventions}) together with a number of ancillary results. In Appendix \ref{app:scalar} we provide a detailed discussion of the classical solutions to wave equations and memories for scalar fields in any $D$, comprising among other things the calculation of null memory induced by a charged particle moving at the speed of light on a test particle initially at rest in odd dimensions. In Appendix \ref{app:box}  we provide a closed-form solution of $\Box \epsilon=0$ in any even dimension with boundary condition $\epsilon(u,r,\mathbf n)\to \lambda(\mathbf n)$ for $r\to\infty$. In Appendix \ref{app:Weinberg} we complete our exploration of the infrared triangle for electromagnetism \cite{Lectures} by describing the link between Weinberg's soft theorem and the asymptotic symmetry in the Lorenz gauge, in even dimensions.

%%%%%Charges, memories and scattering: examples
\section{Classical EM Solutions and Memory Effects} \label{sec:memories}

In this section we calculate the fields generated by specific background charged particles, both massive and massless. 
In particular, we evaluate in any even $D$ the memory effects induced on a test particle at a large distance $r$ from the origin, much in the spirit of \cite{Garfinkle:2017fre, WaldOddD, Mao-Ouyang}, while also including the effects induced by massless charged particles, thus extending the four-dimensional analysis of \cite{BG2013}. Throughout the paper we shall work in the Lorenz gauge, which in Cartesian coordinates reads $\partial^\mu \mathcal A_\mu  =0$, so that the equations of motion reduce to a set of scalar wave equations
\be
\Box\, \mathcal A_\mu = j_\mu\,.
\ee
Thus, much of the calculations are essentially the same as those of the scalar case, that we detail in Appendix \ref{app:scalar}.

 First, we consider the unphysical but instructive case of a static point-like source with charge $q$ created at the origin,\footnote{Strictly speaking, equation \eqref{createdcha} is not well-posed, since the right-hand side has a non-zero divergence. We shall take this aspect in due account when calculating the solution \eqref{dim6vectorpot}, where in particular the right-hand side of \eqref{createdcha} occurs just as part of a full source that respects the continuity equation.} 
\be\label{createdcha}
\Box\, \mathcal A^\mu = u^\mu q\,\theta(t)\delta(\mathbf x)\,,
\ee
where $u^\mu=(1,0,\ldots,0)$. The retarded solution is
\be
\mathcal A^\mu = -\, u^\mu \varphi\,,
\ee
where $\varphi$ denotes the corresponding solution \eqref{scalar-basic} for the scalar field that we report here for completeness
\be\label{scalar-basicBIS}
\varphi(u,r)=q
\sum_{k=0}^{D/2-2} c_{D,k}\,
\frac{\theta^{(D/2-2-k)}(u)}{r^{D/2-1+k}}\,.
\ee
Here $\theta (u)$ denotes the Heaviside distribution, while the coefficients $c_{D,k}$ are specified in \eqref{coefficients_scalar}. The coordinates $u$ and $r$ are part of the retarded Bondi coordinates, in which the Minkowski metric takes the form
\be
ds^2 = - du^2 - 2 du dr + r^2 \g_{ij} dx^i dx^j\,.
\ee
The solution describing a source created at the origin with velocity $\mathbf{v}$ is obtained by boosting the previous one according to \eqref{standardboost} and yields $\mathcal A^\mu=(\mathcal A^0, \mathbf A)=-\,\gamma(\mathbf v)(1,\mathbf v)\varphi$. Moving to retarded coordinates gives
\be
\mathcal A_u = \gamma(\mathbf v)\varphi\,,\qquad
\mathcal A_r = \gamma(\mathbf v)(1-\mathbf n \cdot \mathbf v)\varphi\,,\qquad
\mathcal A_i = -\,r\,\gamma(\mathbf v) v_i\, \varphi\,,
\ee
where $\mathbf n = \mathbf n(x^i)$ is the unit vector identifying points on the $(D-2)$-sphere and $v_i = \mathbf v \cdot \partial_i\mathbf n$.

 Focussing for simplicity on the case of $D=6$, let us now consider the radiation field generated by a massive particle with charge $q$ sitting at rest in the origin for $t<0$, that starts moving with velocity $\mathbf v$ at $t=0$. Such a field is obtained by matching the solution for a charge destroyed at the origin to that of a charge  there created with velocity $\mathbf v$. Proceeding as described in Appendix \ref{app:scalar}, one obtains, for large values of $r$,
\be 
\begin{aligned}\label{dim6vectorpot}
	8\pi^2\mathcal A_u &= 
	\frac{q\,\delta(u)}{r^2}\left(\frac{1}{1-\mathbf n \cdot \mathbf v}-1 \right)
	+\mathcal O(r^{-3})
	\,,\\
	8\pi^2\mathcal A_r &= \mathcal {O}(r^{-3})\,,\\
	8\pi^2\mathcal A_i &= -\frac{q\,v_i\,\delta(u)}{(1-\mathbf n\cdot \mathbf v)\,r}-\frac{q\,v_i\,\theta(u)}{\gamma(\mathbf v)^2\,(1-\mathbf n\cdot \mathbf v)^3\,r^2}+\mathcal {O}(r^{-3})\,,
\end{aligned} 
\ee
where we kept the orders relevant to the calculation of the memory effect.
The change in the angular components of the momentum $P_\mu$ of a test charge $Q$, initially at rest, gives rise to a \emph{linear} (or \emph{ordinary}) memory effect that, to leading order, reads
\be\label{memoryformulaD=6}
P_i\big|_{u>0}-P_i\big|_{u<0} = Q \int_{-\infty}^{+\infty} \mathcal F_{iu}\, du' = \frac{q\,Q\,v_i(2-\mathbf v^2-\mathbf n \cdot \mathbf v)}{8\pi^2\,(1-\mathbf n\cdot \mathbf v)^3\,r^2}+\mathcal {O}(r^{-3})\,,
\ee 
where  $\cF_{\m\n}$ is the Maxwell field strength.

With hindsight, having in mind in particular the results of \cite{Mao-Ouyang}, in order to interpret this leading memory effect in terms of a symmetry, it is useful to perform a further gauge fixing in order to get rid of the residual symmetries affecting the radiation order. Choosing a gauge parameter
\be
8\pi^2\epsilon = -\frac{q\,\theta(u)}{r^2} \left(\frac{1}{1-\mathbf n \cdot \mathbf v}-1 \right)+\mathcal O(r^{-3})
\ee 
allows us to cancel the leading term of $\mathcal A_u$, compatibly with the condition $\Box \epsilon=0$.
The resulting field after performing this gauge transformation satisfies\footnote{More explicitly, one has $\mathcal A_u = \tfrac{q}{8\pi^2r^3}+\mathcal O(r^{-4})$.}
\be
\mathcal A_u = \mathcal O({r^{-3}})\,,\quad
8\pi^2\mathcal A_i= 
-\frac{q\,v_i\,\delta(u)}{(1-\mathbf n \cdot \mathbf v)\,r}
-\frac{q\,v_i\,\theta(u)(2-\mathbf v^2-\mathbf n \cdot \mathbf v)}{(1-\mathbf n \cdot \mathbf v)^3\,r^2}
+\mathcal O(r^{-3})\,,
\ee 
so that, in particular, $\mathcal A_i$ is fully responsible for the memory formula \eqref{memoryformulaD=6}. The effect 
is proportional to the variation of the latter field component between $u>0$ and $u<0$ and takes the form of a total derivative  on the sphere, close to null infinity: 
\be
\begin{aligned}\label{specificexamplememory}
\mathcal A_i\big|_{u>0}-\mathcal A_i\big|_{u<0} 
&= -\frac{q\,v_i\,(2-\mathbf v^2-\mathbf n \cdot \mathbf v)}{(1-\mathbf n \cdot \mathbf v)^3\,r^2}+\mathcal O(r^{-3})\\
&=-\frac{q}{r^2}\,\partial_i\! \left(
\frac{1-\mathbf v^2}{2(1-\mathbf n \cdot \mathbf v)^2}+\frac{1}{1-\mathbf n \cdot \mathbf v}
\right)+\mathcal O(r^{-3})\,.
\end{aligned}
\ee  
This result provides an explicit connection between the memory effect and a residual symmetry acting, for large values of $r$, at Coulombic order.\footnote{The Coulombic order is identified by the condition that the modulus of the vector $\mathcal A_\mu$ in the given metric scales like $\mathcal O(r^{3-D})$. In particular, at this order, $\mathcal A_i\sim \mathcal O(r^{4-D})$.} 

Let us turn our attention to the case of \emph{null} memory. The four-dimensional case was illustrated in \cite{BG2013}. Let us consider, in any even dimension $D\ge6$, the field generated by a charge moving in the ${\mathbf {\hat x}}$ direction at the speed of light: in Cartesian coordinates,
\be
\Box \mathcal A^\mu = q\, v^\mu \delta(\mathbf x- {\mathbf {\hat x}} t)\,,
\ee
with $v^\mu=(1,{\mathbf {\hat x}})$ and $|{\mathbf {\hat x}}|=1$. Taking into account the corresponding retarded solution \eqref{nullmemk} for the scalar field $\varphi$, we then have $A^\mu = - v^\mu \varphi $ and, moving to retarded coordinates,
\be
\mathcal A_u = \varphi\,,\qquad
\mathcal A_r = (1-\mathbf n \cdot {\mathbf {\hat x}})\varphi\,,\qquad
\mathcal A_i = -\,r\, \hat x_i\, \varphi\,.
\ee
Consequently
\be
\begin{aligned}\label{AnullD}
	\mathcal A_u &\sim -\frac{q\,\delta(u)}{r^{D-4}}\,(\Delta-D+4)^{-1}(\mathbf n, {\mathbf {\hat x}})\,,\\
	\mathcal A_r &\sim -\frac{q\,\delta(u)}{r^{D-4}}(1-\mathbf n \cdot \mathbf x_0)\,(\Delta-D+4)^{-1}(\mathbf n, {\mathbf {\hat x}})\,,\\
	\mathcal A_i &\sim \frac{q\,\hat x_i \delta(u)}{r^{D-5}}\,(\Delta-D+4)^{-1}(\mathbf n, {\mathbf {\hat x}})\,,
\end{aligned}
\ee
where $(\Delta-D+4)^{-1}$ denotes the Green function for the operator $\Delta-D+4$ and we have omitted terms of the form
\be
\frac{\delta^{(D-4-k)}(u)}{r^k}\,F(\mathbf n)
\ee 
(see \eqref{nullmemk}), which would be \emph{leading} with respect to those displayed in \eqref{AnullD}, for suitable values of $k$, but which do not contribute to the memory effect because they integrate to zero. The null memory formula then reads
\be\label{nullmemDeven}
P_i\big|_{u>0}-P_i\big|_{u<0}= Q\int_{-\infty}^{+\infty} \mathcal F_{iu}\, du' = -\frac{q\,Q}{r^{D-4}}\,\partial_i(\Delta-D+4)^{-1}(\mathbf n, {\mathbf {\hat x}})\,,
\ee
where this result is exact for the solution considered, while for more general solutions of the wave equation in the form  \eqref{recursionnullD} it would provide the leading-order contribution. 

%%%%%
\section{Electromagnetic Memory}\label{sec:em}
%%%%%

In the previous section we saw how to establish a connection between local symmetries acting at large $r$ and memory effects for specific matter configurations. 
Now we would like to investigate this connection beyond those examples by studying the general structure of the solution space in the Lorenz gauge. To this end, we will analyse the asymptotic behaviour of the fields in a power-series expansion in the radial coordinate in  $D>6$, which is sufficient to the description of the memory effects. More general possibilities, also encompassing the four-dimensional case, will be discussed in Section~\ref{sec:poly}, where we will explore the full structure of the asymptotic symmetries in the Lorenz gauge. 

%%%%%
\subsection{Electromagnetism in the Lorenz gauge}\label{ssec:lorenz}
%%%%%

The Lorenz gauge condition reads 
\be\label{Lorenz}
\nabla^\mu \mathcal A_\mu
=
-\,\partial_u \mathcal A_r + \frac{1}{r} \left( r\partial_r + D-2\right)(\mathcal A_r - \mth A_u) + \frac{1}{r^2}\,\cD^i \mth A_i =0\,,
\ee
where $\cD_i$ is the covariant derivative on the Euclidean unit $(D-2)$-sphere with metric $\g_{ij}$. 
The residual symmetry parameters then satisfy $\Box \epsilon=0$, namely
\be\label{residual}
\left(2\,\partial_r + \frac{D-2}{r}\right)\partial_u \epsilon = \left(\partial_r^2 + \frac{D-2}{r}\,\partial_r + \frac{1}{r^2}\,\Delta\right)\epsilon \,.
\ee
The equations of motion reduce to $\Box \mth A_\mu=0$, which, component by component, read
\begin{subequations} \label{eom}
\begin{align}
&\left[\partial_r^2-2\partial_u\partial_r - 
\frac{D-2}{r} (\partial_u-\partial_r)+\frac{1}{r^2}\Delta\right]\mth A_u=0\, ,\\[10pt]
&\left[\partial_r^2-2\partial_u\partial_r - \frac{D-2}{r}(\partial_u-\partial_r) + \frac{1}{r^2}\Delta \right]\mth A_r+\frac{D-2}{r^2}(\mth A_u - \mth A_r)-\frac{2}{r^3} \cD\cdot \mth A=0\, ,\\[10pt]
&\left[\left(\partial_r^2-2\partial_u\partial_r + \frac{1}{r^2}\Delta\right)-\frac{D-4}{r}(\partial_u-\partial_r)-\frac{D-3}{r^2}\right]\mth A_i-\frac{2}{r}\cD_i(\mth A_u -\mth A_r)=0\,.
\end{align}
\end{subequations}
%

%%%%%
\subsection{Asymptotic expansion}\label{ssec:asympt}
%%%%%

We assume an asymptotic expansion of the gauge field and of the gauge parameters in powers of $1/r$ 
\be
\mathcal A_\mu = \sum_k A_\mu^{(k)} r^{-k}\,,
\qquad
\epsilon = \sum_k \epsilon^{(k)}r^{-k}\,,
\ee
where the summation ranges are, for the moment, unspecified.
Equations \eqref{Lorenz} and \eqref{residual} then give 
\begin{align}
\partial_u A^{(k+1)}_r&=(k-D+2)(A_u^{(k)}-A_r^{(k)})+\cD\cdot A^{(k-1)}
\label{Lorenz_r}\, ,\\[10pt]
(D-2k-2)\partial_u \epsilon^{(k)}&=[\Delta + (k-1)(k-D+2)]\epsilon^{(k-1)}
\label{residual_r}\, , 
\end{align}
while from \eqref{eom} one obtains
\begin{align}
&(D-2k-2)\partial_u A_u^{(k)}=[\Delta + (k-1)(k-D+2)]A_u^{(k-1)}
\label{equ_r}\, ,\\[10pt]
&(D-2k-2)\partial_u A^{(k)}_r = [\Delta+k(k-D+1)]A_r^{(k-1)} + (D-2)\, A_u^{(k-1)}-2 \cD\cdot\! A^{(k-2)}
\label{eqr_r}\, ,\\[10pt]
&(D-2k-4)\partial_u A_i^{(k)}=[\Delta+k(k-D+3)-1]A_i^{(k-1)}-2\cD_i(A_u^{(k)}-A_r^{(k)})
\label{eqi_r}\,.
\end{align}
Eqs. \eqref{Lorenz_r}---\eqref{eqi_r} appear in particular to order $r^{-(k+1)}$ in the asymptotic expansions of the original equations.

As we shall see, to the purpose of analysing the electromagnetic memory effects the leading falloffs can be chosen so as to match the corresponding radiation falloffs: 
\be\label{radiationfall}
\mathcal A_u = \mathcal O(r^{-(D-2)/2})\,,\qquad
\mathcal A_r = \mathcal O(r^{-(D-2)/2})\,,\qquad
\mathcal A_i = \mathcal O(r^{-(D-4)/2})\,.
\ee 
More general options are possible and influence the structure of the asymptotic symmetry group. We will be concerned with these more general aspects of the discussion in Section \ref{sec:poly}.

The significance of the choice \eqref{radiationfall} lies in the fact that the derivatives with respect to $u$ of the field components are unconstrained to leading order: these conditions for the asymptotic expansion are well-suited to identifying the boundary data for a radiation solution with an arbitrary wave-form. Such components also provide the energy flux at a given retarded time, according to
\be
\mathcal P(u) = \int_{S_u}  \gamma^{ij} \partial_u A^{(\frac{D-4}{2})}_i \partial_u A^{(\frac{D-4}{2})}_j d\Omega\,,
\ee
where $S_u$ is the section of $\mathscr I^+$ at fixed $u$ and $d\Omega$ is the measure element on the unit $(D-2)$-sphere.

The asymptotic behaviour of radiation differs in higher dimensions with respect to the characteristic Coulombic falloff $r^{3-D}$ that, in its turn, can be identified as the leading falloff for stationary solutions. (See also the discussion in Section \ref{ssec:em_memory} on this point.) As we shall see,  Coulomb fields give non-vanishing contributions to the surface integral associated with the electric charge as well as to the memory effects.

%%%%%
\subsubsection{Recursive gauge fixing}\label{ssec:emgaugefix}
%%%%%

The gauge variations
\be
\delta A_u^{(k)}=\partial_u \epsilon^{(k)}\,,\qquad
\delta A_r^{(k)}=-(k-1)\epsilon^{(k-1)}\,,\qquad
\delta A_i^{(k)}=\partial_i \epsilon^{(k)}\,
\ee 
imply  a number of restrictions on the allowed gauge parameters, in order to keep the corresponding falloffs. From $\d \cA_r = \cO(r^{-\frac{D-2}{2}})$ we read off $\epsilon^{(k)}=0$ for $k<(D-4)/2$ and $k\neq 0$, while $\delta A_u^{(k)} = \cO(r^{-\frac{D-2}{2}})$ additionally requires that $\epsilon^{(\frac{D-4}{2})}$ be independent of $u$, whereas $\delta A_i^{(k)}$ does not give rise to further constraints at this stage. This leads to a gauge parameter of the following form
\be \label{epsilon}
\epsilon(u,r,x^i)=1+ r^{-\frac{D-4}{2}} \epsilon^{(\frac{D-4}{2})}(x^i) +
\mathcal O(r^{-\frac{D-2}{2}})\,,
\ee
where we conventionally set the global part of $\epsilon$ to $1$.
However, by \eqref{residual_r},
\be
\left[\Delta - \frac{(D-2)(D-4)}{4}\right]\epsilon^{(\frac{D-4}{2})}=0\,,
\ee
which implies $\epsilon^{(\frac{D-4}{2})}=0$ since the Laplacian on the sphere is negative semidefinite.\footnote{
Interestingly, however, let us observe that the corresponding putative charge 
\be \nonumber
\mathcal {\tilde Q}_\epsilon = \lim_{r\to\infty}\int_{S_u} \epsilon\, \mathcal F_{ur} r^{D-2} d\Omega
= \int \cD\cdot A^{(\frac{D-4}{2})} \epsilon^{(\frac{D-4}{2})} d\Omega
\ee
is finite and non-vanishing as $r\to\infty$.
}
Thus, we need to search further down in the asymptotic expansion of $\epsilon$, employing a recursive on-shell gauge-fixing procedure.
The residual symmetry is parametrised as follows
\be
\epsilon(u,r,x^i)=1+ r^{-\frac{D-2}{2}} \epsilon^{(\frac{D-2}{2})}(u,x^i)+\cdots\,.
\ee 
Equation \eqref{residual_r} leaves the $u$-dependence of $\epsilon^{(\frac{D-2}{2})}$ unconstrained, and therefore we may use it to set 
\be
A_u^{(\frac{D-2}{2})}=0\,,
\ee
leaving a residual parameter
$\epsilon^{(\frac{D-2}{2})}(x^i)$ arbitrary.
Setting $k=D/2$, \eqref{equ_r} and $\eqref{residual_r}$ reduce to
\be
\partial_u A_u^{(\frac{D}{2})}=0\,,\qquad
\partial_u \epsilon^{(\frac{D}{2})}+\left[\Delta-\frac{(D-2)(D-4)}{4}\right]\epsilon^{(\frac{D-2}{2})}=0\,,
\ee
respectively. Thus, $A_u^{(\frac{D}{2})}$ is a function of the angles $x^i$ only, while $\delta A_u^{(\frac{D}{2})}=\partial_u \epsilon^{(\frac{D}{2})}$ can be expressed in terms of $\epsilon^{(\frac{D-2}{2})}$, which
can be used to set $A_u^{(\frac{D}{2})}=0$, while still leaving
$\epsilon^{(\frac{D}{2})}(x^i)$ arbitrary.
We proceed recursively, assuming
\be
A_u^{(\frac{D-2}{2})}=A_u^{(\frac{D}{2})}=\cdots=A_u^{(q-1)}=0\,,\qquad \epsilon^{(q-1)}(x^i)\quad\text{arbitrary}\,,
\ee
for some $q>D/2$.
Then, for $k=q-1$, \eqref{equ_r} and $\eqref{residual_r}$ give
\be\label{induction-equations}
\begin{split}
&(D-2q-2)\partial_u A_u^{(q)}=0\,,\\[5pt]
&(D-2q-2)\partial_u\epsilon^{(q)}=\left[\Delta-(q-1)\left(D-q-2\right)\right]\epsilon^{(q-1)}\,.
\end{split}
\ee
Therefore, we may employ $\epsilon^{(q-1)}(x^i)$ to set $A_u^{(q)}$ to zero provided that the differential operator on the right-hand side is invertible, which is true for any $q<D-2$.

We shall now consider two options. We may first choose to truncate the recursive gauge fixing right after the step labelled by $q=D-4$, which leaves us with the asymptotic expansions
\be\label{falloffmemory1}
\mathcal A_u = \sum_{k=D-3}^\infty A_u^{(k)}r^{-k}\,,\qquad
\mathcal A_r = \sum_{k=\frac{D}{2}}^\infty A_r^{(k)}r^{-k}\,,\qquad
\mathcal A_i = \sum_{k=\frac{D-4}{2}}^\infty A_i^{(k)}r^{-k}\,,
\ee
where $A_r^{(\frac{D-2}{2})}=0$ on shell.
The residual symmetry is then given by
\be\label{res_even}
\epsilon(u,r,x^i) \, =\, 1+ \epsilon^{(D-4)}(x^i)r^{4-D}\, + \cdots\,,
\ee
whose corresponding charge, evaluated in the absence of radiation close to the past boundary $\mathscr I^+_-$ of $\mathscr I^+$, reads\footnote{Similar considerations will apply to the evaluation of soft charges in Section \ref{sec:poly}.}
\be\label{ChargeStrom2}
\mathcal Q_\epsilon = \frac{1}{r^{D-4}}\int_{\mathscr I^+_-}
\left(
\partial_u A^{(D-2)}_r+(D-3)A_u^{(D-3)}
\right)
\epsilon^{(D-4)}d\Omega_{D-2}\,.
\ee
Alternatively, we may also perform the recursive gauge-fixing until the very last allowed step, $q=D-3$, thus obtaining
\be\label{falloffmemory2}
\mathcal A_u = \sum_{k=D-2}^\infty A_u^{(k)}r^{-k}\,,\qquad
\mathcal A_r = \sum_{k=\frac{D}{2}}^\infty A_r^{(k)}r^{-k}\,,\qquad
\mathcal A_i = \sum_{k=\frac{D-4}{2}}^\infty A_i^{(k)}r^{-k}
\ee
and
\be
\epsilon(u,r,x^i) \, =\, 1+ \epsilon^{(D-3)}(x^i)r^{3-D}\, + \cdots\,.
\ee
The latter choice highlights the possibility of making the components
\be\label{gaugeinvtower}
A_i^{(\frac{D-4}{2})},\cdots, A_{i}^{(D-5)}, A_i^{(D-4)}
\ee
gauge-invariant, and hence in principle responsible for any observable effect due to radiation impinging on a test charge placed at a large distance $r$ from a source.
Indeed, consistently with the examples of Section \ref{sec:memories}, in the next section we shall see that electromagnetic memory effects appear at the Coulombic order $A_i^{(D-4)}$.

%%%%%
\subsection{Electromagnetic memory}\label{ssec:em_memory}
%%%%%

A test particle with charge $Q$, initially at rest at a large distance $r$ from the origin, will experience a leading-order momentum kick due to the presence of an electric field according to  
\be\label{velocity}
P_{i}\big|_{u_1}-P_i\big|_{u_0} = Q\int_{u_0}^{u_1} \mathcal F_{iu}\, du\,.
\ee 
We are neglecting the contribution from the magnetic field, assuming the interaction time between the test particle and the radiation is sufficiently small. In particular, \eqref{velocity} holds to leading order for an ideally sharp wave front, as we checked in Section~\ref{sec:memories}. 

Now we shall consider solutions that are stationary before $u=u_0$ and after $u=u_1$. For such solutions, the Maxwell tensor does not vanish outside this interval, in general, because static forces are present. However, it does not contain radiation, and thus all the components of the gauge potential associated to radiation are to vanish, or, more generally, are to be pure gauge. In particular, the radiation field components before $u_0$ and after $u_1$ are to be identical or are to differ by a gauge transformation.

Let us combine the information of the previous recursive gauge-fixing with the requirement that the solution be stationary before $u_0$ and after $u_1$.
The coefficients $F_{iu}^{(k)}$ of the asymptotic expansion of the electric field $\mth F_{iu}=\partial_u \mathcal A_i-\partial_i \mth A_u$ satisfy
\be
F_{iu}^{(k)} = -\partial_u A^{(k)}_i(u,x^k)\,,  
\ee
for $k=\frac{D-4}{2},\ldots,D-4$
since the corresponding $A_u^{(k)}$ are zero, thanks to the gauge-fixing. Thus, 
\be\label{DeltaPDeltaA}
P_i\big|_{u_1}-P_i\big|_{u_0} =-Q \sum_{k=\frac{D-4}{2}}^{D-4}\frac{1}{r^k}\left( A^{(k)}_i\big|_{u_1}-A_i^{(k)}\big|_{u_0} \right)+ \mathcal O(r^{3-D})\,.
\ee
With respect to our previous observation, let us notice that the components of $\mathcal A_i$ that enter the subleading terms $\mathcal O(r^{3-D})$ are those connected with non-vanishing stationary properties of the field (we assume no long-range magnetic forces), while all the leading components explicitly written enter the radiation behaviour and thus their difference after $u_1$ and before $u_0$ can be at most the angular gradient of given functions. In addition, we shall immediately see that, combining this information with the equations of motion will allow us to appreciate that they all vanish with the exception of the last one at order  $r^{4-D}$.

Indeed, let us note that for a stationary solution, in our gauge,
equations \eqref{Lorenz_r}, \eqref{equ_r}, \eqref{eqr_r} and \eqref{eqi_r} read, for $k<D-2$, 
\be
(D-k-1)(A_r^{(k-1)}-A_u^{(k-1)})+\cD\cdot A^{(k-2)}=0
\label{Lorenz_ru=0}\,,
\ee
and
\begin{align}
[\Delta + (k-1)(k-D+2)]A_u^{(k-1)}&=0
\label{equ_ru=0}\, ,\\[6pt]
[\Delta+k(k-D+1)]A_r^{(k-1)}\, +\, (D-2)\, A_u^{(k-1)}-2 \cD\cdot A^{(k-2)}&=0
\label{eqr_ru=0}\,,\\[6pt]
[\Delta+(k-1)(k-D+2)-1]A_i^{(k-2)}-2\cD_i(A_u^{(k-1)}-A_r^{(k-1)})&=0
\label{eqi_ru=0}\,.
\end{align}
For $1<k<D-2$, equation \eqref{equ_ru=0} implies $A_u^{(k-1)}=0$ consistently with the recursive gauge-fixing. Equations \eqref{Lorenz_ru=0} and \eqref{eqr_ru=0} then give, for $1<k<D-2$,
\be
[\Delta+(k-2)(k-D+1)]A_r^{(k-1)}=0
\ee
so that $A_r^{(k-1)}=0$ for $2<k<D-2$. Considering finally equation \eqref{eqi_ru=0}, for $2<k<D-2$, we have
\be
[\Delta+(k-1)(k-D+2)-1]A_i^{(k-2)}=0
\ee
and hence $A_i^{(k-2)}=0$ provided provided that $k$ also satisfies $k_-(D)<k<k_+(D)$ with 
\be
k_\pm(D) = \frac{1}{2}\left[D-1\pm \sqrt{(D-3)^2+4} \right]\,;
\ee
actually, $k_-(D)<1$ and $k_+(D)>D-2$ for any $D>3$, so we conclude that stationary solutions obey $A_i^{(k-2)}=0$ for $2<k<D-2$.

To summarise, $A_u^{(k)}=0$ for $0<k<D-3$, while $A_r^{(k)}=0$ for $1<k<D-3$ and 
\be
A_i^{(k)}=0 \quad \text{for} \quad 0<k<D-4\,.
\ee 
By equation \eqref{DeltaPDeltaA}, the condition on $\mathcal A_i$ implies that the leading-order memory effect appears at $\mathcal O(r^{4-D})$,
\be\label{memoryform_even}
P_i\big|_{u_1}-P_i\big|_{u_0} = -\frac{Q}{r^{D-4}}\left( A^{(D-4)}_i\big|_{u_1}-A_i^{(D-4)}\big|_{u_0} \right)+\mathcal O(r^{3-D})\,.
\ee
In view of the discussion below \eqref{DeltaPDeltaA}, we conclude that the momentum shift must take the form 
\be\label{memoryformulaf}
P_i\big|_{u_1}-P_i\big|_{u_0} = \frac{Q}{r^{D-4}}\,\partial_ig(x^k)+\mathcal O(r^{3-D})\,,
\ee
where $g(x^i)$ is a $u$-independent function, which depends on the shape of the radiation train and in particular on $u_0$ and $u_1$ (see for instance the example \eqref{specificexamplememory}). 

Let us note that, as it must be, this difference is not affected by the action of the residual gauge transformation \eqref{res_even}, because the latter is $u$-independent and thus cannot alter the difference $A_i^{(D-4)}\big|_{u_1}-A_i^{(D-4)}\big|_{u_0}$.
In this sense, whether or not one performs the last step of the recursive gauge fixing is irrelevant to the extent of calculating the electromagnetic memory.

To conclude, we have established a formula that, for any even $D$, exhibits a momentum kick characterising the transition between the initial and final vacuum configurations, parametrised by the gauge transformation $g(x^i)$, induced by the exposure to electromagnetic radiation reaching null infinity. In particular, the norm of this effect scales as $r^{3-D}$. 

Up to this point we have only been dealing with an ordinary memory effect; in order to encompass null memory, we must modify the equations of motion \eqref{eom} by adding a suitable source term to the right-hand side, namely a current density $J^\mu$ allowing for the outflow to future null infinity of charged massless particles. The falloff conditions on such a current can be taken as follows
\be
J_u=\mathcal O(r^{2-D})\,,\qquad
J_r=\mathcal O(r^{2-D})\,,\qquad
J_i=\mathcal O(r^{3-D})\,.
\ee
This is clearly displayed by the current for a single massless charge $q$ moving in the ${\mathbf {\hat x}}$ direction, whose components read in Minkowski coordinates
\be
J^0 = q\, \delta(\mathbf x - {\mathbf {\hat x}} t)\,,\qquad
\mathbf J =q\, {\mathbf {\hat x}}\, \delta(\mathbf x - {\mathbf {\hat x}} t)\,,
\ee 
and in retarded coordinates (for $t=u+r>0$)
\be\begin{aligned}
J_u &= - \frac{q}{r^{D-2}}\,\delta(u)\delta(\mathbf n, {\mathbf {\hat x}})\,,\\
J_r &= -\frac{q(1-\mathbf n \cdot {\mathbf {\hat x}})}{r^{D-2}}\,\delta(u)\delta(\mathbf n, {\mathbf {\hat x}})\,,\\
J_i &= \frac{q \hat x_i}{r^{D-3}}\,\delta(u)\delta(\mathbf n, {\mathbf {\hat x}})\,.
\end{aligned}\ee
To the purposes of the recursive gauge fixing, the only modification is thus the introduction of a source term $J^{(D-2)}_u$ in the right-hand side of \eqref{equ_r} when $k=D-3$, which now actually forces us to stop the gauge fixing after the use of the said equation for $k=D-4$ (the step labelled by $q=D-4$ in the previous section) and leaves us with the falloff \eqref{falloffmemory1}. On the contrary, reaching \eqref{falloffmemory2} is not allowed, and thus \eqref{falloffmemory1} comprises a complete gauge fixing. However, also in view of the above considerations, the discussion of the memory effect and its relation to the symmetry acting at Coulombic order remain unaltered.

\subsection{Phase memory}\label{ssec:phasemem}

Let us consider a pair of electric charges $q$ and $-q$ that are pinned in the positions $(r,\mathbf n_1)$ and $(r,\mathbf n_2)$, for large $r$. We will now derive the expression for an imprint that the passage of a radiation train leaves on the properties of these particles that is encoded in the phase of their states. 
What follows is inspired by the four-dimensional discussion of \cite{Susskind:2015hpa}. For a quantum treatment of electromagnetic kick memory see instead \cite{ShahinString}.

To this purpose, let us assume that, as in the previous section, radiation impinges on the charges only during the interval between two given retarded times $u_0$ and $u_1$. As we have seen, this means that
\be\label{gaugevacua}
\mathcal A_i\big|_{u_1}-\mathcal A_i\big|_{u_0}=\frac{1}{r^{D-4}}\,\partial_i g + \mathcal O(r^{3-D})
\ee
for a suitable angular function $g(\mathbf n)$. We will assume for simplicity that the gauge field before the onset of radiation is the trivial one.

Let $|\psi_1\rangle=|q\rangle$ and $|\psi_2\rangle=|-q\rangle$ be the initial states in which the charged particles are prepared, which are uniquely labelled by their charges since translational degrees of freedom have been suppressed. Before $u_0$, the state $|\psi_2,\mathbf n_1\rangle$ obtained by the parallel transport of the second state $|\psi_2\rangle$ to the position $\mathbf n_1$ of the first the charge is 
\be
|\psi_2,\mathbf n_1\rangle=|\psi_2\rangle\,,
\ee 
because $\mathcal A_\mu=0$, so that the corresponding tensor state evaluated in $\mathbf n_1$ is given by 
\be
|\psi_1,\mathbf n_1\rangle \otimes |\psi_2,\mathbf n_1\rangle=|\psi_1\rangle\otimes |\psi_2\rangle\,.
\ee 

At any fixed $u\ge u_1$, instead, the same operation of parallel transport must be performed by calculating
\be
|\psi_2,\mathbf n_1\rangle = 
\exp\left[iq\int_{\mathbf n_2}^{\mathbf n_1}\mathcal A_i \,dx^i\right]|\psi_2\rangle = 
\exp\left[iq\,\frac{g(\mathbf n_1)-g(\mathbf n_2)}{r^{D-4}}\right]|\psi_2\rangle+\mathcal O(r^{3-D})\,,
\ee
where we have employed \eqref{gaugevacua}.
Therefore, after the passage of radiation,
\be
|\psi_1, \mathbf n_1\rangle \otimes |\psi_2,\mathbf n_1\rangle = \exp\left[iq\,\frac{g(\mathbf n_1)-g(\mathbf n_2)}{r^{D-4}}\right]|\psi_1\rangle\otimes |\psi_2\rangle+\mathcal O(r^{3-D})\,,
\ee
which displays how the transition between two different radiative vacua, already experimentally detectable by the occurrence of a non-trivial velocity kick for a test charge, is also signalled by the variation of the relative phases in the states obtained by parallel transport of charged particles. Such a phase can be non-trivial provided that the function $g$ is non-constant, namely when there is a non-trivial memory kick \eqref{memoryformulaf}. 

A point that should be stressed is that, in this setup, the states $|\psi_1\rangle$ and $|\psi_2\rangle$ do not evolve, since each particle is kept fixed in its position (its translational quantum numbers are \emph{frozen}) while electromagnetic radiation cannot change its charge. The relative phase difference occurs entirely as an effect of the evolution of $\mathcal A_\mu$, which undergoes a transition between two underlying radiative vacua. As we shall see in the following, this aspect is qualitatively different in a non-Abelian theory such as Yang-Mills, where radiation can alter the colour charge.

%%%%%
\section{Yang-Mills Memory} \label{sec:color}
%%%%%

In this section we extend the previous analysis to the non-Abelian case.

%%%%%
\subsection{Yang-Mills theory in the Lorenz gauge}\label{ssec:lorenzYM}
%%%%%
We consider pure Yang-Mills theory of an anti-Hermitian gauge field $\mathcal A_\mu(x)=\mathcal A_\mu^A(x) T^A$, where $T^A$ are the generators of an $su(N)$ algebra. 
The field strength reads $\mathcal F_{\mu\nu} = \partial_\mu \mathcal A_\nu - \partial_\nu \mathcal A_\mu + [\mathcal A_\mu, \mathcal A_\nu]$, while the infinitesimal form of the gauge transformation is expressed by $\delta_\epsilon \mathcal A_\mu = \partial_\mu \epsilon + [\mathcal A_\mu, \epsilon]$ in terms of the gauge parameter $\epsilon(x) = \epsilon^A(x) T^A$.

We impose the Lorenz gauge condition $\nabla^\mu \mathcal A_\mu=0$,
which leaves as residual gauge parameters those that satisfy
$
\Box \epsilon +[\mathcal A_\mu, \partial^\mu \epsilon]=0
$.
Furthermore, the equations of motion reduce to $\Box \mathcal A_\nu + [\mathcal A^\mu, \nabla_\mu A_\nu+\mathcal F_{\mu\nu}]=0$.

Adopting retarded Bondi coordinates, the Lorenz gauge condition is tantamount to
\be\label{Lorenz_YM}
\partial_u \mathcal A_r = \left(\partial_r + \frac{D-2}{r}\right)(\mathcal A_r - \mth A_u)+\frac{1}{r^2}\cD\cdot \mth A\,,
\ee
while the constraint on residual transformations is
\be \label{residual_YM}
\left(\partial_u^2 - 2 \partial_u \partial_r + \frac{1}{r^2} \Delta \right) \epsilon + \frac{D-2}{r}(\partial_r - \partial_u)\epsilon =
[\partial_r \epsilon, \mathcal A_r - \mathcal A_u] - [\partial_u \epsilon, \mathcal A_r] + \frac{1}{r^2}\gamma^{ij}[\cD_i \epsilon, \mathcal A_j]\,.
\ee
The equations of motion are instead
\beal[]\label{eom_YM}
&\left[\partial_r^2-2\partial_u\partial_r - 
\frac{D-2}{r} (\partial_u-\partial_r)+\frac{1}{r^2}\Delta\right]\mth A_u\\
&=
[\mth A_u-\mth A_r, \partial_r \mth A_u + \mth F_{ru}]+[\mth A_r, \partial_u \mth A_u]-\frac{\gamma^{ij}}{r^2}[\mth A_i, \cD_j\mth A_u + \mth F_{ju}]\, ,
\\[10pt]
&\left[\partial_r^2-2\partial_u\partial_r \, - \, \frac{D-2}{r}(\partial_u-\partial_r)\, + \, \frac{1}{r^2}\Delta \right]\mth A_r+\frac{D-2}{r^2}(\mth A_u - \mth A_r)-\frac{2}{r^3} \cD\cdot \mth A\\
&=
[\mth A_u- \mth A_r , \partial_r \mth A_r] + [\mth A_r, \partial_u \mth A_r + \mth F_{ur}] - \frac{\gamma^{ij}}{r^2}[\mth A_i, \cD_j \mth A_r+ \mth F_{jr}]
\, ,
\\[10pt]
&\left[\left(\partial_r^2-2\partial_u\partial_r + \frac{1}{r^2}\Delta\right)-\frac{D-4}{r}(\partial_u-\partial_r)-\frac{D-3}{r^2}\right]\mth A_i-\frac{2}{r}\cD_i(\mth A_u -\mth A_r)\\
&=
\left[\mth A_u- \mth A_r, \left(\partial_r - \frac{2}{r}\right)\mth A_i + \mth F_{ri}\right] + [\mth A_r, \partial_u \mth A_i + \mth F_{ui}]-\frac{\gamma^{jk}}{r^2}[\mth A_j, \cD_k\mth A_i+\mth F_{ki}]\,.
\eeal
%

%%%%%
\subsection{Asymptotic expansion and colour flux}\label{ssec:asymptYM}
%%%%%
Performing the usual asymptotic expansion in inverse powers of the radial coordinate $r$, one obtains the following set of equations.
Equations \eqref{Lorenz_YM} and \eqref{residual_YM} give
\be
\partial_u A^{(k+1)}_r=(D-k-2)(A_r^{(k)}-A_u^{(k)})+\cD\cdot A^{(k-1)}
\label{Lorenz_r_YM}\, 
\ee
and
\beal
&(D-2k-2)\partial_u \epsilon^{(k)}-[\Delta - (k-1)(D-k-2)]\epsilon^{(k-1)}\\ 
&= \sum_{l+m=k}
\left(
-l[\epsilon^{(l)}, A_u^{(m)}-A_r^{(m)}]+[\partial_u \epsilon^{(l+1)}, A_r^{(m)}]-\gamma^{ij}[\cD_i \epsilon^{(l-1)}, A_j^{(m)}]
\right)
\label{residual_r_YM}\, ,
\eeal 
respectively, while from \eqref{eom_YM} one obtains
\begin{align}
\label{equ_r_YM}
&(D-2k-2)\partial_u A_u^{(k)}-[\Delta - (D-k-2)(k-1)]A_u^{(k-1)}
\\ \nonumber
= & \sum_{l+m=k}
\Big(
[A_r^{(m)}- A_u^{(m)}, -l A_u^{(l)}+ F_{ru}^{(l+1)}]-[A_r^{(m)}, \partial_u A_u^{(l+1)}]
\\ \nonumber
&+\gamma^{ij}[A_i^{(m)}, \cD_j A_u^{(l-1)} + F_{ju}^{(l-1)}]
\Big)
\, ,\\[10pt]
\label{eqr_r_YM}
&(D-2k-2)\partial_u A^{(k)}_r \!-\! [\Delta-k(D-k-1)]A_r^{(k-1)}\! -\! (D-2) A_u^{(k-1)}+2 \cD\cdot A^{(k-2)}
\\ \nonumber
= & \sum_{l+m=k}
\Big([A_r^{(m)}-A_u^{(m)}, -l A_r^{(l)}]-[A_r^{(m)}, \partial_u A_r^{(l+1)}+F_{ur}^{(l+1)}]
\\ \nonumber
&+\gamma^{ij}[A_i^{(m)}, \cD_j A_r^{(l-1)}+F_{jr}^{(l-1)}]
\Big)
\, ,\\[10pt]
\label{eqi_r_YM}
&(D-2k-4)\partial_u A_i^{(k)}-[\Delta-k(D-k-3)-1]A_i^{(k-1)}+2\cD_i(A_u^{(k)}-A_r^{(k)})
\\ \nonumber
= & \sum_{l+m=k}
\Big([A_r^{(m)}-A_u^{(m)}, -(l+2) A_i^{(l)}+F_{ri}^{(l+1)}]-[A_r^{(m)}, \partial_u A_i^{(l+1)}+F_{ui}^{(l+1)}]
\\ \nonumber
&+\gamma^{jk}[A_j^{(m)}, \cD_k A_i^{(l-1)}+F_{ki}^{(l-1)}]
\Big)
\, ,
\end{align}
for the corresponding components of the equations of motion. Eqs. \eqref{Lorenz_r_YM}---\eqref{eqi_r_YM} appear to order $r^{-(k+1)}$ in the asymptotic expansions. 

We choose to adopt the same ``radiation'' falloff conditions \eqref{radiationfall} that we imposed in the linearised theory: the asymptotic expansions of $\mth A_u$ and $\mth A_r$ start at order $r^{-(D-2)/2}$ and that of $\mth A_i$ starts at order $r^{-(D-4)/2}$.

For completeness, and as a cross-check of the choice of falloffs, let us now discuss the definition of the total colour charge $\mth Q(u) = \mth Q^A(u) T^A$ and its dependence on the retarded time $u$. In the Lorenz gauge, this quantity is given by the surface integral 
\beal
\mathcal Q^A(u) 
=& \int_{S_u} F_{ur}^A r^{D-2}d\Omega\\
=& \sum_k r^{D-2-k} \int_{S_u} \Big(\partial_u A_r^{(k)} + (k-1) A_u^{(k-1)} + \sum_{l+m=k}[A_u^{(m)}, A_r^{(l)}] \Big)^A d\Omega\,,
\eeal 
in the limit $r\to\infty$.
Combining the Lorenz condition \eqref{Lorenz_r_YM} and the $r$ component of the equations of motion \eqref{eqr_r_YM}, one obtains
\beal \label{finite-charge-instr_YM}
&(D-2-k)\left(
\partial_u A_r^{(k)}+(k-1)A_u^{(k)}
\right)
-
\cD^i \left(
\cD_i A_r^{(k-1)}+(k-2)A_i^{(k-2)}
\right)\\
=&
\sum_{l+m=k}\Big(
[A_r^{(m)}-A_u^{(m)}, -l A_r^{(l)}]-[A_r^{(m)}, \partial_u A_r^{(l+1)}+F_{ur}^{(l+1)}]\\
&
+\gamma^{ij}[A_i^{(m)}, D_j A_r^{(l-1)}+F_{jr}^{(l-1)}]
\Big)
\,.
\eeal
We see that \eqref{Lorenz_r_YM} implies $\partial_u A^{(\frac{D-2}{2})}_r=0$ and that \eqref{finite-charge-instr_YM} reduces to 
\be
(D-2-k)\left(
\partial_u A_r^{(k)}+(k-1)A_u^{(k-1)}
\right)
=
\cD^i \left(
\cD_i A_r^{(k-1)}+(k-2)A_i^{(k-2)}
\right)
\ee
for $k<D-2$. 
The final expression for the colour charge may be cast in the form 
\be
\mth Q^A(u) = \int_{S_u} \Big( 
A_r^{(D-3)}
+(D-4)A_u^{(D-3)}
+[A_u^{(\frac{D-2}{2})}, A_r^{(\frac{D-2}{2})}]
\Big)^A d\Omega\,,
\ee
by means of the Lorenz condition \eqref{Lorenz_r_YM}. 

Concerning the dependence of $\mth Q^A$ on retarded time,
recalling that $\partial_u A_r^{(\frac{D-2}{2})}=0$ by the Lorenz condition,
and employing \eqref{equ_r_YM}, \eqref{finite-charge-instr_YM} for $k=D-3$, we get
\be\label{GeneralColorEvolution}
\frac{d}{du}\mathcal Q^A(u) = \int_{S_u} \gamma^{ij} [A_i^{(\frac{D-4}{2})}, \partial_u A_j^{(\frac{D-4}{2})}] d\Omega\,.
\ee 
This formula provides the flux of the total colour charge due to non-linearities of the theory. Note in particular that  the right-hand side involves the radiation components, representing the flux of classical gluons across null infinity.

%%%%%
%\subsection{Recursive gauge fixing}\label{ssec:YMgaugefix}
%%%%%

%%%%%
\subsection{Yang-Mills memory}\label{ssec:YM_memory}
%%%%%

Starting from the radiation falloffs \eqref{radiationfall}, we may employ the residual gauge symmetry of the theory to perform a further, recursive gauge fixing. Actually, since the non-linear corrections to equations \eqref{induction-equations} appear to order $q=D-3$ or higher, the discussion of this gauge fixing proceeds as in Section \ref{ssec:emgaugefix} up to the next to last step. Indeed, in accordance with the fact that Yang-Mills theory must encompass both ordinary/linear and null/non-linear memory, the gauge fixing stops at $q=D-4$ and cannot be performed up until $q=D-3$, as already observed in the case of null electromagnetic memory.

The resulting falloffs after this procedure are thus
\be
\mathcal A_u = \sum_{k=D-3}^\infty A_u^{(k)}r^{-k}\,,\qquad
\mathcal A_r = \sum_{k=\frac{D-2}{2}}^\infty A_r^{(k)}r^{-k}\,,\qquad
\mathcal A_i = \sum_{k=\frac{D-4}{2}}^\infty A_i^{(k)}r^{-k}\,,
\ee
with residual symmetry parameter
\be\label{res_even_YM}
\epsilon(u,r,x^i) \, =\, c^A T^A+ \epsilon^{(D-4)}(x^i)r^{4-D}\, + \cdots\,,
\ee
where $c^A$ are constant coefficients.

A coloured test particle with charge $Q = Q^A T^A$ interacts with the background Yang-Mills field by the Wong equations \cite{WongEquations} 
\be\label{Wong}
\dot P^\mu = \mathrm{tr} (Q \mth F^{\mu\nu}) \dot x_\nu\,,\qquad
\dot Q = -\dot x^\mu[\mathcal A_\mu, Q] \,.
\ee 
Focussing on the region near null infinity, a test ``quark'', initially at rest, invested by radiation between the retarded times $u_0$ and $u_1$, will therefore experience the following leading-order momentum kick  
\be\label{velocity_YM}
P'_{i}-P_i = 
\int_{u_0}^{u_1} \mathrm{tr}(Q \mathcal F_{iu}) du 
=
-\frac{1}{r^{D-4}}\mathrm{tr}(Q {A_i}^{(D-4)})\Big|_{u_1}
=
-\frac{1}{r^{D-4}}\mathrm{tr} [Q\, e^{-\epsilon^{(D-4)}}\partial_i e^{\epsilon^{(D-4)}}]\,,
\ee
where we have chosen the vacuum configuration at $u=u_0$ to be $\mathcal A_\mu =0$.
On the other hand, the colour of the test quark will change, to leading order, according to 
\be
Q'-Q = - \int_{u_0}^{u_1} [\mathcal A_u, Q]\,du = - \frac{1}{r^{D-3}} \int_{u_0}^{u_1}[A_u^{(D-3)},Q]\,du\,,
\ee
where the $u$-dependence of $\mathcal A_u$ is governed, according to equation \eqref{eqr_r_YM} with $k=D-3$, by
\be\label{ColorDelta}
\partial_u A_u^{(D-3)} = \frac{1}{D-4}\, \gamma^{ij}[A_i^{(\frac{D-4}{2})}, \partial_u A_j^{(\frac{D-4}{2})}]\,.
\ee
Equation \eqref{ColorDelta} characterises the evolution of the charge of the quark in terms of the leading outgoing radiation terms.

In order to better understand the dependence of this momentum kick and of colour evolution on the incoming radiation near null infinity, let us further analyse the equations of motion. Combining equation \eqref{Lorenz_r_YM} and \eqref{eqr_r_YM} we see that
\be\label{recrad_1}
\partial_u A_r^{(k+1)} = \mathscr D_k A_r^{(k)}
\ee
for $k<D-3$, where 
\be\label{operatoresfera}
\mathscr D_k=[D+(k-1)(k-D+2)]/(D-2k-2)\,,
\ee 
and 
\be\label{recrad_2}
\partial_u A_r^{(D-2)} = \mathscr D_{D-3} A_r^{(D-3)} - A_u^{(D-3)} - \frac{1}{D-4}\mth J\,,
\ee
for $k=D-3$ in dimensions $D>4$, where we have defined
\be
\mathcal J = 
 2 \gamma^{ij} [A_i^{(\frac{D-4}{2})}, \cD_j A_r^{(\frac{D-2}{2})}]
-2 [A_r^{(\frac{D-2}{2})}, \partial_u A_r^{(\frac{D}{2})}]\,.
\ee 
Equations \eqref{Lorenz_r_YM} and \eqref{eqr_r_YM}, evaluated for $k=D-3$ and $k=D-2$, respectively, actually imply the following constraint on $\mth J$,
\be \label{constr_J}
\mth J = -\cD^i [\cD_i A_r^{(D-3)}+(D-4)A_i^{(D-4)}]\,,
\ee
and, taking the derivative of the previous equation with respect to $u$,
\be\label{Coul-rad_1}
(D-4) \partial_u \cD\cdot A^{(D-4)}=-\Delta \partial_u A_r^{(D-3)}
+2 [A_r^{(\frac{D-2}{2})}, \partial_u^2 A_r^{(\frac{D}{2})}] - 2\gamma^{ij}[\partial_u A_i^{(\frac{D-4}{2})}, \cD_j A_r^{(\frac{D-2}{2})}]\,.
\ee
Starting from eq. \eqref{recrad_2} and employing \eqref{recrad_1} recursively, we find that
\be
\partial_u^{(\frac{D}{2})}A_r^{(D-2)}
=
\prod_{l=\frac{D-2}{2}}^{D-3}\mathscr D_l \partial_u A_r^{(\frac{D-2}{2})}
-\partial_u^{\frac{D-2}{2}} A_u^{(D-3)} - \frac{1}{D-4} \partial_u^{\frac{D-2}{2}} \mth J\,,
\ee
and, by equation \eqref{constr_J},
\be\label{Coul-rad_2}
\partial_u^{\frac{D-2}{2}}\cD\cdot A^{(D-4)} = \partial_u^{\frac{D-2}{2}} (\partial_u A_r^{(D-2)}+A_u^{(D-3)})
-\frac{1}{D-4} \partial_u^{\frac{D-2}{2}} \Delta A_r^{(D-3)} - \prod_{l=\frac{D-2}{2}}^{D-3} \mathscr D_l \partial_u A_r^{(\frac{D-2}{2})}\,.
\ee
Equations \eqref{Coul-rad_1} and \eqref{Coul-rad_2} encode the dependence of $\cD\cdot A^{(D-4)}$, and hence of the memory kick, on the radiation data encoded in this case by $A_r^{(\frac{D-2}{2})}$. In dimension $D=4$, these equations take the simpler form
\be
\partial_u \cD^i[A_i^{(0)}, A_r^{(1)}]=0\,,
\qquad
\partial_u \cD\cdot A^{(0)} = \partial_u (\partial_u A_r^{(2)}+A_u^{(1)}) = \partial_u F_{ur}^{(2)}\,.
\ee

\subsection{Colour memory} \label{ssec:colourmem}

Let us now consider a pair of colour charges that are pinned in the positions $(r,\mathbf n_1)$ and $(r,\mathbf n_2)$, for large $r$. 
Analysing the effect of the passage of Yang-Mills radiation  will provide the non-Abelian analogue of the phase memory effect highlighted in Section \ref{ssec:phasemem}, namely a relative rotation between their states in colour space. The content of this section is basically inspired by \cite{StromingerC}, where colour memory was first discussed and whose four-dimensional results we extend to arbitrary even-dimensional spaces.  

We consider coloured particles in the fundamental representation of the gauge group that, for definiteness, we first take to be $SU(2)$. Assuming that radiation is non-trivial only between two given retarded times $u_0$ and $u_1$, as we have seen, the angular components of the gauge field satisfy
\be\label{gaugevacua1}
\mathcal A_i\big|_{u_1}-\mathcal A_i\big|_{u_0}=\frac{1}{r^{D-4}}\,e^{-\alpha}\partial_i e^{\alpha}+ \mathcal O(r^{3-D})
\ee
for a suitable $\alpha(\mathbf n)=\alpha^A(\mathbf n)T^A$, where $T^A$ are the $su(2)$ generators. We have assumed for simplicity that the pure gauge configuration before $u_0$ is the trivial one.

Let $|\psi_1\rangle$ and $|\psi_2\rangle$ be the initial states in which the coloured particles are prepared. Before $u_0$, the state $|\psi_2,\mathbf n_1\rangle$ that results from the parallel transport of the second state $|\psi_2\rangle$ to the position $\mathbf n_1$ of the first the charge is
\be
|\psi_2,\mathbf n_1\rangle=|\psi_2\rangle\,,
\ee 
because $\mathcal A_\mu=0$. In particular, one can take it to be one of the eigenstates of $T^3$, \emph{i.e.}\ $|+\rangle$ or $|-\rangle$. In order to build a colour singlet state in $\mathbf n_1$ it then suffices to prepare the superposition 
\be\label{prepare-singlet}
\frac{|+,\mathbf n_1\rangle \otimes |-,\mathbf n_1\rangle-|-,\mathbf n_1\rangle \otimes |+,\mathbf n_1\rangle}{\sqrt2} =\frac{|+\rangle\otimes|-\rangle-|-\rangle\otimes|+\rangle}{\sqrt2}\,.
\ee 
We remark that, in order for the concept of singlet to be well-defined when $SU(2)$ transformations can depend on the position, it is crucial to build a singlet out of states that have been parallel-transported to the same point.

A given state $|\psi\rangle$ evolves according to the covariant conservation equation \cite{LibroRusso}\footnote{The covariant conservation equation for the colour states, which in general reads $\left(\tfrac{d}{d\tau}+\dot x^\mu \mathcal A_\mu\right)|\psi\rangle=0$, implies the Wong equation \eqref{Wong} for the colour charge $\tfrac{d}{d\tau}Q+[\dot x^\mu \mathcal A_\mu,Q]=0$, where $Q=Q^AT^A$ and $Q^A=\langle \psi|T^A|\psi\rangle$.}
\be
\partial_u |\psi\rangle = - \mathcal A_u |\psi\rangle
\ee
where we recall that $\mathcal A_u = \mathcal O(r^{3-D})$ for large $r$. Thus, the state after the passage of radiation differs with respect to its initial value by
\be\label{colour-evolution}
|\psi\rangle\big|_{u_1}-|\psi\rangle \big|_{u_0}=\mathcal O(r^{3-D})\,.
\ee

Now, at $u_1$ one must also take into account the effect of parallel transport from $\mathbf n_2$ to $\mathbf n_1$ on the sphere,  given by
\be\begin{aligned}
|\psi_2,\mathbf n_1\rangle &= 
\mathcal P\exp\left[-\int_{\mathbf n_2}^{\mathbf n_1}\mathcal A_i \,dx^i\right]|\psi_2\rangle \\
&= 
|\psi_2\rangle
-
\frac{1}{r^{D-4}}\int_{\mathbf n_2}^{\mathbf n_1}e^{-\alpha}\partial_i e^{\alpha}dx^i\,|\psi_2\rangle+\mathcal O(r^{3-D})\,,
\end{aligned}\ee
where $\mathcal P$ denotes path ordering and we have employed \eqref{gaugevacua1} to establish the second equality.
We conclude that the actual state  after radiation has passed is no longer the singlet \eqref{prepare-singlet}, but rather:
\be\begin{aligned}\label{falsesinglet}
&\frac{|+,\mathbf n_1\rangle \otimes |-,\mathbf n_1\rangle-|-,\mathbf n_1\rangle \otimes |+,\mathbf n_1\rangle}{\sqrt2} 
= 
\frac{|+\rangle\otimes|-\rangle-|-\rangle\otimes|+\rangle}{\sqrt2}\\
&-
\frac{1}{r^{D-4}\sqrt2}
\left[
|+\rangle\otimes 
\int_{\mathbf n_2}^{\mathbf n_1}e^{-\alpha}\partial_i e^{\alpha}dx^i\,|-\rangle
-
|-\rangle\otimes
\int_{\mathbf n_2}^{\mathbf n_1}e^{-\alpha}\partial_i e^{\alpha}dx^i\,|+\rangle
\right]+\mathcal O(r^{3-D})
\,.
\end{aligned}\ee

Comparison between  \eqref{prepare-singlet} and \eqref{falsesinglet} shows that the interaction of the colour charges with the external Yang-Mills radiation induces a rotation of the initial state that manifests itself  to order $\mathcal O(r^{4-D})$ in the final state. In other words, a pair of particles initially prepared in a singlet will no longer be in a singlet after the passage of radiation. 
This colour memory can be non-trivial provided that the function $\alpha$ is non-constant, namely whenever there is also potentially a non-trivial memory kick \eqref{memoryformulaf}. 

From a technical point of view, it should be noted how the effect of radiation on each colour state is to induce a time evolution to order $\mathcal O(r^{3-D})$, according to \eqref{colour-evolution}, which allows one to disentangle it from the leading effect due to the vacuum transition \eqref{gaugevacua1} undergone by $\mathcal A_i$, which enters \eqref{falsesinglet} to order $\mathcal O(r^{4-D})$. 

A very similar derivation allows one to extend the result to more general states and gauge groups. Adopting an orthonormal basis $|n\rangle$ for the fundamental representation of $SU(N)$, we can consider the superposition
\be\label{beforenm}
\sum_{n,m} c_{n,m}|n\rangle\otimes|m\rangle
\ee
with $\sum_{n,m}|c_{n,m}|^2=1$,
prepared before $u_0$. Time evolution will induce modifications of $|n\rangle$ that appear to order $\mathcal O(r^{3-D})$ by $\mathcal A_u=\mathcal O(r^{3-D})$. On the other hand, the effect of parallel transport from $\mathbf n_2$ to $\mathbf n_1$ performed after $u_1$ gives rise to the leading-order correction 
\be
-\frac{1}{r^{D-4}}\sum_{n,m} c_{n,m}|n\rangle\otimes \int_{\mathbf n_2}^{\mathbf n_1}e^{-\alpha}\partial_i e^{\alpha} dx^i\, | m\rangle
\ee
to \eqref{beforenm},
due to the non-trivial configuration \eqref{gaugevacua1} attained by $\mathcal A_i$.

\section{More on Asymptotic Symmetries for the Maxwell Theory}
\label{sec:poly}

So far, we focussed on the symmetries directly related to memory effects. As we saw, these symmetries act at Coulombic order so that the corresponding charge \eqref{ChargeStrom2} displays a falloff $\mathcal O(r^{4-D})$. Thus, they do not comprise asymptotic symmetries, \emph{stricto sensu}, since the associated charges vanish at $\mathscr I^+$.\ft{In addition, one should check whether the symmetry is actually canonically generated by those charges. This type of analysis, however, lies beyond the scope of our work.} In this section we aim to complement our discussion  by performing a more general analysis of the asymptotic symmetry algebra for the Maxwell theory in higher dimensions.\ft{Let us mention, however, that contrary to the calculation of memory effects, which appear to be tightly related to the properties of the wave equation in even-dimensional spacetimes, these further investigations of asymptotic symmetries admit an extension to spacetimes of not only  even but also odd dimension. While focussing on the former case in the main discussion, we will comment along the way on the differences arising in the latter.}

In the radial gauge, in $D=4$, the standard power-like asymptotic expansion easily allows one to identify the existence of infinite-dimensional asymptotic symmetries. The same result, in the same gauge, can be obtained in higher dimensions upon choosing a power-like ansatz with leading-order terms as weak as $1/r$ \cite{StromingerQEDevenD}. (Showing the consistency of this choice in the gravitational case allowed to identify the BMS group as the relevant group of asymptotic symmetries for asymptotically flat spacetimes of arbitrary even dimension \cite{alternative-bnd}.) We would like to perform the same analysis in the Lorenz gauge. 

Differently from the radial gauge, in the Lorenz gauge one first observes that a mere power-like ansatz with leading-order falloffs as weak as $1/r$ does not allow to recover a non-trivial structure for the asymptotic symmetry group. In order to display the latter, one finds the need for generalising the asymptotic expansion by allowing for logarithmic terms. In this section we will provide an explicit realisation to this effect, while also discussing the non-obvious issue of the 
finiteness of the corresponding soft surface charges.

The observed difference between radial and Lorenz gauge is one concrete facet of a general issue concerning the possible gauge-dependence of the asymptotic analysis \cite{Campiglia-Soni,CD2form,MarcCedric-anyD,Riello}. While in some instances, such as the present one, this problem has been clarified, the general systematics may still deserve further attention.

%%%%%
\subsection{Polyhomogeneous expansion for $D\ge4$}
%%%%%

In the Lorenz gauge, in order to investigate the possible existence of large gauge symmetries acting at $\mathscr I^+$ we look for solutions to equation \eqref{residual} with the boundary condition
\be\label{boundaryfuncI}
\lim_{r\to\infty}\epsilon(u,r,\mathbf n) = \epsilon^{(0)}(u,\mathbf n)\,,
\ee 
for some non-constant function $\epsilon^{(0)}(u,\mathbf n)$. As it turns out, a power-law ansatz
\begin{equation}
\epsilon = \sum_{k=0}^\infty \frac{\epsilon^{(k)}(u,\mathbf n)}{r^k}\,,
\end{equation}
effectively selects only the global symmetry in any even dimension $D\ge4$. Indeed, considering \eqref{residual_r} with $k=0,1,\ldots,\frac{D-2}{2}$, we have
\beal\label{no-symmetry-even}
(D-2)\partial_u \epsilon^{(0)}&=0\,,\\
(D-4)\partial_u \epsilon^{(1)}&=\Delta\epsilon^{(0)}\,,\\
&\,\,\,\vdots\\
2\partial_u\epsilon^{(\frac{D-4}{2})}&=\left[\Delta-\frac{D(D-6)}{4}\right]\epsilon^{(\frac{D-6}{4})}\,,\\
0&=\left[\Delta-\frac{(D-4)(D-2)}{4}\right]\epsilon^{(\frac{D-4}{2})}\,.
\eeal
The last equation sets $\epsilon^{(\frac{D-4}{2})}$ to zero for any $D\ge6$, and to a constant for $D=4$. Then, the above system recursively sets to zero the other components $\epsilon^{(k)}$ all the way up to $\epsilon^{(0)}$ since the corresponding operators are all invertible. Eventually, $\epsilon^{(0)}$ must be a constant by $\partial_u\epsilon^{(0)}=0$ and $\Delta \epsilon^{(0)}=0$.

One concludes that the power-law ansatz does not allow for an enhanced asymptotic symmetry sitting at the order $\cO(1)$. In order to retrieve it, one needs a more general ansatz. We find that it is sufficient, to this purpose, to consider the following type of asymptotic expansion\footnote{
	The fully polyhomogeneous ansatz would be of the form $\sum_{k,j} r^{-k} (\log r)^j \epsilon^{(k,j)}(u,\mathbf n)$, involving arbitrary powers of $\log r$. We do not explore this more general possibility in the present paper. 
} involving also a logarithmic dependence on $r$ \cite{Hirai-Sugishita, MitraHePhoton, MitraHeGauge, MarcCedric-anyD}:
\be\label{expepsilonlog}
\epsilon(u,r,\mathbf n)=
\sum_{k=0}^\infty \frac{\epsilon^{(k)}(u,\mathbf n)}{r^k}
+
\sum_{k=\frac{D-2}{2}}^\infty\frac{\lambda^{(k)}(u,\mathbf n)\log r}{r^k}\,.
\ee
In this fashion, the last equation in \eqref{no-symmetry-even} becomes modified by the presence of the logarithmic branch and yields
\be\label{matching-logpot}
2\partial_u \lambda^{(\frac{D-2}{2})}=\left[\Delta-\frac{(D-4)(D-2)}{4}\right]\epsilon^{(\frac{D-4}{2})}\,,
\ee 
an equation that determines the $u$-dependence of $\lambda^{(\frac{D-2}{2})}$, thus still allowing for an arbitrary $\epsilon^{(0)}(\mathbf n)$. 

Indeed, the recursion relations expressing the equation $\Box\epsilon=0$ read
\beal\label{expansioneq-with-logs}
(D-2-2k)\partial_u \epsilon^{(k)}
+
2\partial_u \lambda^{(k)}
&=
[\Delta +(k-1)(k-D+2)]\epsilon^{(k-1)}
+
(D-1-2k)\lambda^{(k-1)}\,,\\[10pt]
(D-2-2k)\partial_u \lambda^{(k)}
&=
[\Delta +(k-1)(k-D+2)]\lambda^{(k-1)}\,,
\eeal
and hence can be solved by direct integration with respect to $u$. Explicitly, for $k<\frac{D-2}{2}$, they reduce to the familiar system
\be
(D-2-2k)\partial_u \epsilon^{(k)}=[\Delta+(k-1)(k-D+2)]\epsilon^{(k-1)},
\ee  
so that the solution is given by a polynomial in $u$ with angle-dependent coefficients $C_{j,k}(\mathbf n)$, with $0\le j\le k$, and is uniquely determined by specifying the integration functions $\hat \epsilon^{(k)}(\mathbf n)$:
\be
\epsilon^{(k)}(u,\mathbf n)=\sum_{j=0}^k C_{j,k}(\mathbf n) u^j\,,
\ee
with  
\be
C_{j,k}(\mathbf n) = 
\begin{cases}
	\hat \epsilon^{(k)}(\mathbf n) &\text{if }j=0\\
	\frac{1}{j!}\prod_{l=k-j+1}^k \mathscr D_l \hat \epsilon^{k-l}(\mathbf n)&\text{otherwise}\,,
\end{cases}
\ee
where $\mathscr D_l$ was defined in \eqref{operatoresfera}.
If $k=\frac{D-2}{2}$, equation \eqref{expansioneq-with-logs} reduces to \eqref{matching-logpot}, so that
\be
\lambda^{(\frac{D-2}{2})}(u,\mathbf n)=\frac{1}{2}\int_0^u\left[\Delta-\frac{(D-4)(D-2)}{4}\right]\epsilon^{(\frac{D-4}{2})}(u',\mathbf n)\, du'+\hat \lambda^{(\frac{D-2}{2})}(\mathbf n) \,,
\ee
with a suitable integration function, while $\epsilon^{(\frac{D-2}{2})}(u,\mathbf n)$ stays unconstrained. For $k>\frac{D-2}{2}$, we can recast \eqref{expansioneq-with-logs} in the slightly more suggestive form
\beal
(D-2-2k)\partial_u \lambda^{(k)}
&=
[\Delta +(k-1)(k-D+2)]\lambda^{(k-1)}\,,\\[10pt]
(D-2-2k)\partial_u \epsilon^{(k)}
&=
[\Delta +(k-1)(k-D+2)]\epsilon^{(k-1)}
\\
&+\left[
\frac{2}{2k-D+2}(\Delta+(k-1)(k-D+2))+(D-1-2k)
\right]\lambda^{(k-1)}\,.
\eeal
In this way, it becomes clear that these two equations can always be solved by first finding the integral of the first equation directly, and then substituting it in the second one where it acts as a ``source'' term on the right-hand side. 

To summarise, a solution of \eqref{expansioneq-with-logs} is specified by assigning a set of integration functions $\hat \epsilon^{(k)}(\mathbf n)$, for $k\ge0$, and $\hat \lambda^{(k)}(\mathbf n)$, for $k\ge\frac{D-2}{2}$, together with an arbitrary function $\epsilon^{(\frac{D-2}{2})}(u,\mathbf n)$. In particular, this achieves the boundary condition \eqref{boundaryfuncI}
as $r\to\infty$, where $\epsilon^{(0)}(\mathbf n)$ is an arbitrary function on the celestial sphere.

The situation is simpler in odd dimensions, where the recursion relation \eqref{residual_r} can be solved compatibly with the boundary condition \eqref{boundaryfuncI} without the need of introducing logarithmic terms. The technical reason is that, when $D$ is odd, the left-hand side of \eqref{residual_r} never vanishes for integer $k$.

By means of the transformations
\be\label{exact...}
\mathcal A_\mu \mapsto \mathcal A_\mu + \partial_\mu \epsilon
\ee 
parametrised by the gauge parameters thus obtained, we can then act on solutions to Maxwell's equations in the Lorenz gauge characterised by the radiation falloffs \eqref{radiationfall} and generate a wider solution space. In particular, in view of the expansion \eqref{expepsilonlog} of $\e$, this type of solutions will generically exhibit the following asymptotic behaviour as $r\to\infty$:
\be\label{dimindep-fall}
\mathcal A_u = \mathcal O(r^{-1})\,,\qquad
\mathcal A_r = \mathcal O(r^{-2})\,,\qquad
\mathcal A_i = \mathcal O(r^{0})\,,
\ee 
in any $D \geq 5$, while
\be\label{diminde4}
\mathcal A_u = \mathcal O(r^{-1}\log r)\,,\qquad
\mathcal A_r = \mathcal O(r^{-1})\,,\qquad
\mathcal A_i = \mathcal O(r^{0})\,,
\ee 
in dimension $D=4$.

It is important to remark that, since such solutions differ from those characterised by \eqref{radiationfall} by a pure gradient, they will still retain all the corresponding physical properties, such as the finiteness of the energy flux at any given retarded time. 

As far as \eqref{exact...} alone is concerned one could in principle consider parameters $\epsilon$ that obey arbitrary asymptotics near $\mathscr I^+$. However, the ones of interest to us are those that define asymptotic symmetries, \emph{i.e.}\ parameters that give rise to finite and nonvanishing asymptotic surface charges. In the next section  we shall see that this is indeed the case for our class of solutions identified by \eqref{boundaryfuncI}.  

\subsection{Asymptotic charges}\label{ssec:Finitecharges}
As it is usually the case in the presence of radiation, the naive surface charge associated to the symmetry \eqref{expepsilonlog}, namely
\be
\mathcal Q_\epsilon(u) = \lim_{r\to\infty} \oint_{S_u}\mathcal F_{ur}(u,r,\mathbf n)\epsilon(u,r,\mathbf n)r^{D-2}\,d\Omega(\mathbf n)\,,
\ee
is formally ill-defined, because the right-hand side contains terms of the type \emph{e.g.}
\be
r^{\frac{D-4}{2}}\oint_{S_u}F_{ur}^{(\frac{D}{2})}\epsilon^{(0)}(\mathbf n)d\Omega\,,
\ee
which do not vanish, even after imposing the equations of motion, precisely due to the presence of an \emph{arbitrary} parameter $\epsilon^{(0)}(\mathbf n)$.

Such difficulties are absent in the case of the global charges because the equations of motion always ensure that these potentially dangerous terms are actually zero. Indeed, $\nabla_\mu \cF^{\mu\nu}=0$ away from electric charges, so that $\int_{S_{u,r}}\cF_{\mu\nu}\,dx^{\mu\nu}$ cannot depend on $r$, nor in fact on $u$, because it must be independent of the specific $(D-2)$-surface under consideration.

It should be noted however that, for the general variation with parameter \eqref{expepsilonlog}, the above difficulties arise only if one attempts to calculate the surface charge by integrating over a sphere at a given retarded time $u$ and radius $r$ and then lets $r$ tend to infinity. On the other hand, the calculation of the charge on a Cauchy surface still gives a well-defined result \cite{MarcCedric-anyD}.
For instance, under the simplifying assumption that the electromagnetic field due to radiation vanish for $u<u_0$ in a neighbourhood of future null infinity, the calculation of the surface charge indeed yields
\be \label{Qu_good}
\mathcal Q_\epsilon(u) 
= \oint_{S_u} F_{ur}^{(D-2)}\epsilon^{(0)}\,d\Omega
=\oint_{S_u}
\big[A_r^{(D-3)}+(D-4)A_u^{(D-3)}\big]\epsilon^{(0)}\,d\Omega\,,
\ee
for \emph{fixed} $u<u_0$, since all radiation components $F_{ur}^{(k)}$, for $k<D-2$, vanish for a stationary solution (see the discussion in Section \ref{ssec:em_memory}) and $S_u$ is indeed the boundary of a Cauchy surface.
Letting then $u$ approach $-\infty$, one has 
\be
\mathcal Q_\epsilon (-\infty)
= \oint_{\mathscr I^+_-} F_{ur}^{(D-2)}\epsilon^{(0)}\,d\Omega
=\oint_{\mathscr I^{+}_-}
\big[ A_r^{(D-3)}+(D-4)A_u^{(D-3)}\big]\epsilon^{(0)}d\Omega\,.
\ee
For $u<u_0$, the quantity $\mathcal Q_\epsilon(u)$ must match the analogous surface integral calculated at spatial infinity because in both cases the Noether two-form is integrated over the boundary of a Cauchy surface, in view of the requirement that no radiation be present in a neighbourhood of $\mathscr I^+$ for $u<u_0$.

In force of the considerations above, we now define the charge $Q_\e(u)$ for all the values of $u$ as follows. For $u < u_0$ we define it as in \eqref{Qu_good}, while for $u \geq u_0$, even in the presence of radiation, we define it as the evolution of \eqref{Qu_good} under the equations of motion. Indeed, the Maxwell equation $\nabla\cdot \mathcal F_r=0$ gives
\be
\Big(\partial_r + \frac{D-2}{r}\Big) \mathcal F_{ur} = \frac{1}{r^2}\cD\cdot \mathcal F_r\,,
\ee
while $\nabla\cdot \mathcal F_u=0$ reads
\be
\partial_u \mathcal F_{ur}=\Big(\partial_r + \frac{D-2}{r}\Big) \mathcal F_{ur}+\frac{1}{r^2}\cD\cdot \mathcal F_u 
\implies
\partial_u \mathcal F_{ur} = \frac{1}{r^2}\cD\cdot(\mathcal F_u+ \mathcal F_r)\,,
\ee 
and hence
\be
\frac{d}{du}\,\mathcal Q_\epsilon (u) = \oint_{S_u} \cD\cdot(F_u^{(D-4)}-F_r^{(D-4)})\epsilon^{(0)}\,d\Omega\,.
\ee
Analogous considerations allow one to introduce well-defined surface charges evaluated at $\mathscr I^-$.

From the perspective of the analysis performed in Section \ref{sec:memories}, it is possible to explicitly calculate the soft charges according to the above strategy. Restricting to the case of a massive charge in dimension $6$ that starts moving at $t=0$, the integral of $\mathcal F_{ur} \epsilon^{(0)}$ on a sphere at fixed retarded time $u$ and radius $r$ yields 
\be\begin{aligned}
	\mathcal Q_\epsilon(r,u) &= 
	r\, \frac{\delta(u)}{8\pi^2}\oint \frac{\mathbf n \cdot \mathbf v(3 \mathbf n \cdot \mathbf v - 4)-\mathbf v^2}{(1-\mathbf n \cdot \mathbf v)^2}\,\epsilon^{(0)}(\mathbf n) \,d\Omega\\
	&+\frac{3}{8\pi^2}\oint \left[\theta(-u)+\theta(u)\frac{(1-\mathbf v^2)^2}{(1-\mathbf n \cdot \mathbf v)^4}\right]\,\epsilon^{(0)}(\mathbf n) \,d\Omega\,.
\end{aligned}\ee
Except for the case of the electric charge $\epsilon^{(0)}=1$, where the first integral vanishes identically, the limit of this surface charge as $r\to\infty$ is ill-defined in the presence of radiation, namely on the forward light-cone $u=0$, due to the linear divergence appearing in the first line. However, the charge is well-defined on $\mathscr I^+$ before and after the passage of radiation, $u\neq0$, and reads
\be
\mathcal Q_\epsilon(u)=\frac{3}{8\pi^2}\oint \left[\theta(-u)+\theta(u)\frac{(1-\mathbf v^2)^2}{(1-\mathbf n \cdot \mathbf v)^4}\right]\,\epsilon^{(0)}(\mathbf n) \,d\Omega\,.
\ee
For $\epsilon^{(0)}=0$, this quantity reduces to the (constant) electric charge $Q=1$, while for more general parameters $\epsilon^{(0)}$, the soft charge exhibits a jump discontinuity at $u=0$, which measures the fact that the particle is no longer static for $u>0$ in a manner akin to the memory effect itself.

Performing instead the limit $r\to\infty$ at fixed time $t$, it is also possible to verify the matching between the surface charge evaluated at null infinity before the onset of radiation, for $u<0$ (or, equivalently, at $\mathscr I^+_-$), and the Hamiltonian charge $\mathcal H_\epsilon(t)$, obtained by integrating on a slice at fixed time $t$. Indeed, taking \eqref{exactscalarD=6} into account and writing the result in terms of polar coordinates $t$, $r$ and $\mathbf n$, we have, for the scalar field,
\be\begin{aligned}\label{exactscalarD=6polar}
	8\pi^2 \varphi &= \frac{\delta(t-r)}{\gamma(\mathbf v)(1-\mathbf n\cdot\mathbf v)r^2}+\frac{\theta(t-r)}{\gamma(\mathbf v)^3(1-\mathbf n\cdot\mathbf v)^3r^3}\,\Delta(t-r,r)^{-3/2}\\
	&-\frac{\delta(t-r)}{r^2}+\frac{\theta(r-t)}{r^3}\,.
\end{aligned}\ee 
The corresponding electromagnetic potential is given by $\mathcal A^\mu=(\mathcal A^0,\mathbf A)=-\gamma(\mathbf v)(1,\mathbf v)\varphi$, for $t>r$, and $A^\mu=(\varphi,0)$, for $t<r$. The radial component of the electric field then yields $\mathcal F_{tr}=3 r^{-4}$ as $r\to\infty$ for fixed $t$
and hence
\be
\mathcal H_\epsilon (t)= \frac{3}{8\pi^2}\oint \epsilon^{(0)}(\mathbf n) \,d\Omega = \mathcal Q_\epsilon(u<0)\,.
\ee

Similar arguments showing the finiteness of the Hamiltonian charge in higher-dimensions have been given, in the case of linearised spin-two case in retarded Bondi gauge, in \cite{Aggarwal}, while a renormalisation procedure has been recently proposed, for the Maxwell theory in the radial gauge, in \cite{FreidelHopfmuellerRiello} (for the general definition of surface charges see \cite{Barnich-Brandt}). The relation between asymptotic charges and the soft photon theorem in any $D$ was recently clarified in \cite{MitraHePhoton, MitraHeGauge}. In Appendix \ref{app:Weinberg} we present the details of the computation for the case of interest in our work.

%%%%%
\acknowledgments
%%%%%
We are grateful to Glenn Barnich and C\'edric Troessaert for discussions and exchanges. A.C. and D.F. express their gratitude  to the Erwin Schr\"odinger International Institute for Mathematics and Physics (ESI) of Vienna for the kind hospitality extended to them during the preparation of this work, in the occasion of the workshop  ``Higher spins and holography''. The work of A.C.\ was partially supported by the Fund for Scientific Research--FNRS PDR grant ``Fundamental issues in extended gravity'', No.~T.0022.19. The work of C.H.\ was partially supported by the Istituto Nazionale di Fisica Nucleare (Iniziativa Specifica GSS-Pi) and by the Fund for Scientific Research--FNRS, Belgium; he also acknowledges the hospitality of ESI. A.C.\ and C.H.\ gratefully acknowledge the University Roma Tre for hospitality.

\appendix
\section{Notation}\label{app:conventions}

Retarded Bondi coordinates are a retarded time $u=t-r$, a radial coordinate $r$, and angular coordinates $x^i$ on the Euclidean unit $(D-2)-$sphere, with metric $\gamma_{ij}$. 
The Minkowski metric, in such coordinates, reads
\be \label{metric-retarded}
ds^2 = -du^2-2 du dr + r^2 \gamma_{ij} dx^i dx^j\,,
\ee
while the (nonvanishing) Christoffel symbols are
\be
\Gamma\indices{^i_{rj}}=\frac{1}{r}\delta\indices{^i_j}\, ,\qquad
\Gamma\indices{^u_{ij}}=-\Gamma\indices{^r_{ij}}=r\gamma_{ij}\, ,\qquad
\Gamma\indices{^i_{jk}}=\frac{1}{2}\gamma^{il}(\partial_j \gamma_{lk}+\partial_k \gamma_{jl}-\partial_l \gamma_{jk})\,.
\ee
With $\cD_i$ we denote the covariant derivative associated to $\gamma_{ij}$ and $\Delta= \cD_i\cD^i$ is the corresponding Laplace-Beltrami operator. In particular, the d'Alembert operator $\Box$ acting on a scalar $\varphi$ takes the explicit form
\be
\Box \varphi  = -\left(2\partial_r + \frac{D-2}{r}\right)\partial_u \varphi + \left(\partial_r^2 + \frac{D-2}{r}\,\partial_r + \frac{1}{r^2}\Delta\right)\varphi\,.
\ee

We find it useful to also employ the notation $\mathbf n = \mathbf n(x^i)$ for unit vectors identifying points on the sphere, in terms of which $\gamma_{ij}=\partial_i\mathbf n \cdot \partial_j\mathbf n$.

\section{Classical Scalar Solutions and Memory Effects}\label{app:scalar}

In this appendix, we derive explicit solutions of the scalar wave equation with different types of background sources and calculate the associated memory effects. This provides a useful warming-up for our discussion of electromagnetic memory effects in the  Lorenz gauge in even dimensions, while also allowing us to shed some light on the nature of memory effects in odd dimensions.

\subsection{Scalar fields in even $D$}\label{sec:spin0}
Let us first consider a particle with charge $q$ under the scalar field $\varphi$ that is created in the origin at $t=0$. The field generated by this process is obtained by solving the wave equation 
\be\label{scalar-theta}
-\Box \varphi(t,\mathbf x) = q\, \theta(t)\delta(\mathbf x)
\ee 
(we adopt the convention $-\Box=-\eta^{\mu\nu}\partial_\mu\partial_\nu=\partial_t^2-\nabla^2$).
The solution is given by the convolution of the source on the right-hand side with the ($D$--dimensional) retarded wave propagator $G^\text{ret}_D(x)$, \emph{i.e.} in this case
\be
\varphi(t,\mathbf x) 
= q\, \int_{-\infty}^{t} G^\text{ret}_D(\tau, \mathbf x)\,d\tau\,.
\ee
The field generated by a particle that is destroyed in the origin can then be obtained by time reversal of the above solution, while the field for a moving particle can be calculated by applying a Lorentz boost. 

Let us recall that \cite{Friedlander}, for even $D\ge2$, the retarded propagator is 
\be\label{prop-even}
G^\text{ret}_D(x)= 
\dfrac{1}{2\pi^{D/2-1}}\delta^{(\frac{D-4}{2})}(x^2)\theta(x^0)\,,
\ee
where $\theta$ is the Heaviside distribution.
Restricting to even $D\ge4$, and using the chain rule for the distribution $\delta^{(\frac{D-4}{2})}(x^2)$, the propagator \eqref{prop-even} can be recast as 
\be
G^\text{ret}_D(t, \mathbf x)= 
\sum_{k=0}^{D/2-2} c_{D,k}\,
\frac{\delta^{(D/2-2-k)}(u)}{r^{D/2-1+k}}\,,
\ee
where $u=t-r$ and $r=|\mathbf x|$, while $c_{D,k}$ are the coefficients
\be \label{coefficients_scalar}
c_{D,k} = \frac{1}{2(2\pi)^{\frac{D}{2}-1}}\,\frac{\left(\frac{D}{2}-2+k\right)!}{2^k\left(\frac{D}{2}-2-k\right)!\,k!}\,.
\ee 
In particular note that
\be
c_{D,0}=\frac{1}{2(2\pi)^{\frac{D}{2}-1}}\,,\qquad
c_{D,\frac{D}{2}-3} = c_{D,\frac{D}{2}-2}= \frac{1}{(D-3)\Omega_{D-2}}\,,
\ee
where $\Omega_{D-2}$ is the area of the $(D-2)$-dimensional Euclidean unit sphere.
The resulting scalar field is thus
\be\label{scalar-basic}
\varphi(u,r)=q
\sum_{k=0}^{D/2-2} c_{D,k}\,
\frac{\theta^{(D/2-2-k)}(u)}{r^{D/2-1+k}}\,.
\ee
Notice that only the term associated to $k=D/2-2$ gives rise to a persistent field for fixed $r$, while the other terms have support localised at $u=0$, namely on the future-directed light-cone with vertex at the particle's creation. This is a general consequence of the recursion relation \eqref{residual_r}, namely 
\be
(D-2k-2)\partial_u \varphi^{(k)}
=
[\Delta+(k-1)(k-D+2)]\varphi^{(k-1)}\,,
\ee 
obeyed by $\varphi=\sum_k \varphi^{(k)}r^{-k}$ near future null infinity, which requires $[\Delta +k(k-D+3)]\varphi^{(k)}=0$ for $0<k<D-3$, and hence $\varphi^{(k)}=0$, for any stationary solution.

Now, a test particle with charge $Q$, held in place at a distance $r$ from the origin, will be subject to a force $f_\mu =Q\,\partial_\mu \varphi(u,r)$ at a given retarded time $u$ due to the presence of the scalar field. Hence, its $D$--momentum $P_\mu$ will in general be subject to the leading-order variation
\be
P_\mu\big|_{u}-P_\mu\big|_{u=-\infty}=Q\int_{-\infty}^u \partial_\mu \varphi(u',r) du'\,.
\ee
For this very simple example, this quantity can be calculated explicitly for any even $D$. The variations of $P_u$ and $ P_r$ in particular yield
\be\label{scalarmem0_u}
P_u\big|_{u>0}-P_u\big|_{u<0}= 
Q\int_{-\infty}^{+\infty} \partial_u\varphi(u', r)\,du' = 
\frac{Qq}{(D-3)\Omega_{D-2}r^{D-3}}\,\qquad
\text{for }u>0\,,
\ee
and
\be\label{scalarmem0_r}
P_r\big|_{u>0}-P_r\big|_{u<0}=
Q\int_{-\infty}^{+\infty} \partial_r\varphi(u', r)\,du'=
-\frac{(D-4)Qq}{(D-3)\Omega_{D-2}r^{D-3}}\,.
\ee
Equation \eqref{scalarmem0_u} simply expresses the fact that the test particle will start feeling the Coulombic interaction energy with the newly created particle in the origin as soon as it crosses the light-cone subtended by the origin of spacetime. On the other hand, \eqref{scalarmem0_r} tells us that the test particle will feel an instantaneous, radial momentum kick, for even dimensions greater than four. Since this process is spherically symmetric, the variations of the angular components $P_i$ vanish identically.

The field emitted by a particle destroyed in the origin at $t=0$ is obtained by sending $u\mapsto -u$ in \eqref{scalar-basic}. The case of a particle moving with velocity $\mathbf v$ can be instead obtained by boosting \eqref{scalar-basic}: 
\be\label{standardboost}
t \mapsto \gamma(\mathbf v)(t-\mathbf v \cdot \mathbf x)\,,\qquad
\mathbf x \mapsto \mathbf x + \mathbf v (\gamma(\mathbf v)-1)\frac{\mathbf v \cdot \mathbf x}{\mathbf v^2}- \gamma(\mathbf v)\mathbf v t\,,
\ee
which gives, for large $r$, denoting $\mathbf n = \mathbf x/r$, 
\be\label{transf-ur}
u \mapsto u\, \gamma(\mathbf v)^{-1}(1-\mathbf n \cdot \mathbf v)^{-1}+\mathcal O(r^{-1})\,,\qquad
r \mapsto r\, \gamma(\mathbf v)(1-\mathbf n \cdot \mathbf v)+\mathcal O(1)\,.
\ee
We can then cast the boosted solution in the following form:
\be\label{boosted-scalar}
\varphi(u,r, \mathbf n) = \,\frac{q\,\theta(u)}{(D-3)\Omega_{D-2}[\gamma(\mathbf v)(1-\mathbf n \cdot \mathbf v)r]^{D-3}} 
+\bar \varphi(u, r, \mathbf n)+\mathcal O(r^{2-D})\,,
\ee
where $\bar\varphi$ is a sum of terms of the type
\be
f_\alpha(r, \mathbf n)\delta^{(\alpha)}(u)\,,\qquad \text{ with }\alpha\ge0\,,
\ee
namely, whose support is localised on the light-cone.
Let us stress that the terms in $\bar \varphi$ formally dominate the asymptotic expansion of $\varphi$ as $r\to\infty$. However, these terms will not contribute to the leading $u$-component of the momentum kick due to the presence of $\delta(u)$ and its derivatives. We can therefore conclude that
\be
P_u\big|_{u>0}-P_u\big|_{u<0} =Q\int_{-\infty}^{+\infty} \partial_u \varphi\,du'= \frac{qQ}{(D-3)\Omega_{D-2}[\gamma(\mathbf v)(1-\mathbf n \cdot \mathbf v)r]^{D-3}}\,,
\ee
in any even $D$. This is, not surprisingly, just the analogue of equation \eqref{scalarmem0_u} for the Coulombic energy in which one needs to account for the relativistic length contraction.

For a more general scattering process involving a number of ``in''  and ``out'' particles destroyed or created in the origin, the result is obtained by linearly superposing solutions and therefore reads ($\eta_a=-1$ for an incoming particle and $\eta_a=+1$ for an outgoing one)
\be
P_u\big|_{u>0}-P_u\big|_{u<0} =
\sum_{a\in\text{in/out}}\frac{\eta_a\,q_a Q}{(D-3)\Omega_{D-2}[\gamma(\mathbf v_a)(1-\mathbf n \cdot \mathbf v_a)r]^{D-3}}\,.
\ee

Calculating radial and angular components of $P_\mu$ requires more effort, since they arise instead from the terms proportional to  $\delta(u)$ whose number increases with the spacetime dimension. They have been given for any even dimension in \cite{Mao-Ouyang} in terms of derivatives of a generating function. For our present, illustrative, purposes, it suffices to consider the first relevant case $D=6$, where the exact solution in the case of the particle created in the origin with velocity $\mathbf v$ is given by 
\be\label{exactscalarD=6}
8\pi^2 \varphi = \frac{\delta(u)}{\gamma(\mathbf v)(1-\mathbf n\cdot\mathbf v)r^2}+\frac{\theta(u)}{\gamma(\mathbf v)^3(1-\mathbf n\cdot\mathbf v)^3r^3}\,\Delta(u,r)^{-3/2} 
\ee
with
\be
\Delta(u,r) = 1
+\frac{2u(\mathbf v^2-\mathbf n \cdot \mathbf v)}{r(1-\mathbf n\cdot\mathbf v)^2}
+\frac{u^2\mathbf v^2}{r^2(1-\mathbf n\cdot\mathbf v)^2}\,.
\ee
The corresponding radial and angular memory effects in $D=6$ are then, to leading order,
\be\begin{split}
P_r\big|_{u>0}-P_r\big|_{u<0} &= \frac{-2Qq}{8\pi^2\gamma(\mathbf v)(1-\mathbf n \cdot \mathbf v)r^3} \,,\\[5pt]
P_i\big|_{u>0}-P_i\big|_{u<0} &= \frac{v_i Qq}{8\pi^2 \gamma(\mathbf v)(1-\mathbf n\cdot \mathbf v)r^2}\,,
\end{split}
\ee
where $v_i = \partial_i \mathbf n \cdot \mathbf v$ is the component of the particle's velocity in the $i$-th angular direction.

While the above examples illustrate the phenomenon of ordinary memory, associated with the field emitted to massive charges that move in the bulk of the spacetime, we can also consider the wave equation with a source term characterising the presence of a massless charged particle, moving along a given direction ${\mathbf {\hat x}}$:
\be
-\Box \varphi = q\, \delta(\mathbf x - {\mathbf {\hat x}} t)\,,
\ee 
with $|{\mathbf {\hat x}}|=1$. This equation can be conveniently solved for any even $D\ge6$ by going to retarded coordinates, where it reads
\be
\left(2\partial_r + \frac{D-2}{r}\right)\partial_u \varphi = \left(\partial_r^2 + \frac{D-2}{r}\partial_r + \frac{1}{r^2}\Delta\right)\varphi+\frac{q}{r^{D-2}}\,\delta(u)\,\delta(\mathbf n, {\mathbf {\hat x}})
\ee
and performing the usual asymptotic expansion $\varphi(u,r,\mathbf n)= \sum \varphi^{(k)}(u, \mathbf n)r^{-k}$, which gives
\be\label{recursionnullD}
(D-2k-2)\partial_u \varphi^{(k)} = [\Delta + (k-1)(k-D+2)]\varphi^{(k-1)}+\delta_{k,D-3}\,\delta(u)\,\delta(\mathbf n, {\mathbf {\hat x}})\,.
\ee
The latter equation is solved by setting $\varphi^{(k)}=0$ for $k\le \frac{D}{2}-2$ and for $k\ge D-3$, while, for $\frac{D}{2}-1\le k \le D-4$,
\be\label{nullmemk}
\varphi^{(k)}(u, \mathbf n)=\delta^{(D-4-k)}(u)C_k(\mathbf n)\,,
\ee
where the functions $C_k(\mathbf n)$ are determined recursively by
\be\begin{aligned}
	(D-2k-2) C_k(\mathbf n) &= [\Delta + (k-1)(k-D+2)]C_{k-1}(\mathbf n)\,,\\[10pt]
	C_{D-4}(\mathbf n) &= -(\Delta-D+4)^{-1}(\mathbf n,{\mathbf {\hat x}})\,.
\end{aligned}\ee
Here, $(\Delta-D+4)^{-1}$ is the Green's function for the operator $\Delta-D+4$, which is unique for $D>4$. As a consequence, the field gives rise to the null memory effect
\be
P_i\big|_{u>0}-P_i\big|_{u<0} = \int_{-\infty}^{+\infty} \partial_i \varphi\, du = -\frac{1}{r^{D-4}}\partial_i(\Delta-D+4)^{-1}(\mathbf n,{\mathbf {\hat x}})
\ee
(note that only the term with $k=D-4$ contributes),
consisting in a kick along a direction tangent to the celestial sphere.

\subsection{Comments on the odd-dimensional case}
In odd dimensions $D\ge3$ the retarded propagator is given by \cite{Friedlander}
\be
G_D(x)=
c\,
(-x^2)_+^{1-\frac{D}{2}}\theta(x^0)\,,
\ee
where
$
c^{-1}=2\pi^{\frac{D}{2}-1}\Gamma(2-\frac{D}{2})
$,
while $(\kappa)_+^\alpha$ is the distribution defined as
\be
\langle 
(\kappa)_+^\alpha, \chi(\kappa)
\rangle =
\int_0^\infty \kappa^\alpha \chi(\kappa)\,d\kappa\,\qquad \text{ for }\alpha>-1\,,
\ee
$\chi(\kappa)$ denoting a generic test function, and analytically continued to any $\alpha\neq -1$, $-2$, $-3$, $\ldots$ by 
\be
\langle 
(\kappa)_+^\alpha, \chi(\kappa)
\rangle =
\frac{(-1)^n}{(\alpha+1)(\alpha+2)\cdots (\alpha+n)}\,
\langle
(\kappa)_+^{\alpha+n}, \chi^{(n)}(\kappa)
\rangle\,
\quad
\text{for }n>-1-\alpha\,.
\ee
A relevant feature of the wave propagator in odd dimensional spacetimes is that its support is not localised on the  light-cone $|t|=r$, in contrast with the case of even dimensions, as it is non-zero also for $|t|>r$. This is to be interpreted as the fact that even an ideally sharp perturbation, $\delta(t,\mathbf x)$, will not give rise to an ideally sharp wave-front, but rather the induced radiation will display a dispersion phenomenon and non-trivial disturbances will linger on even after the first wave-front has passed.

The solution to equation \eqref{scalar-theta} is then furnished by 
\beal
\varphi(t,\mathbf x) = 
\frac{cq}{2}
\left\langle 
\kappa^{1-D/2}_+,
\theta(t-\sqrt{\kappa+|\mathbf x|^2})/\sqrt{\kappa+|\mathbf x|^2}
\right\rangle\,.
\eeal
Integrating by parts, and assuming $t>r=|\mathbf x|$ (otherwise the field vanishes by causality), one obtains the following expansion
\be\begin{aligned}\label{phioddD}
	\frac{2}{cq}\,\varphi &= 
	\frac{(t^2-r^2)^{2-\frac{D}{2}}}{(2-\frac{D}{2})t}
	+
	\frac{(t^2-r^2)^{3-\frac{D}{2}}}{(2-\frac{D}{2})(3-\frac{D}{2})t^3}\,\tfrac{1}{2}
	+
	\frac{(t^2-r^2)^{4-\frac{D}{2}}}{(2-\frac{D}{2})(3-\frac{D}{2})(4-\frac{D}{2})t^5}\,\tfrac{1}{2}\cdot\tfrac{3}{2}
	+\cdots\\
	&+\frac{(t^2-r^2)^{-\frac{1}{2}}}{(2-\frac{D}{2})(3-\frac{D}{2})\cdots (-\frac{3}{2})(-\frac{1}{2})t^{D-4}}\,\tfrac{1}{2}\cdot\tfrac{3}{2}\cdots \left(\tfrac{D}{2}-3\right)\\
	&+(-1)^{\frac{D-3}{2}}\int_0^{t^2-r^2}\frac{d\kappa}{\sqrt\kappa (\kappa+r^2)^{\frac{D}{2}-1}}\,.
\end{aligned}\ee 
Moving to retarded coordinates, this result can be recast as
\be\label{scalar-odd}
\varphi(u,r) = 
\bar\varphi(u,r)
+\frac{cq}{2}
(-1)^{\frac{D-3}{2}}\theta(u)
\int_0^{u(u+2r)}
\frac{d\kappa}{\sqrt{\kappa}(\kappa+r^2)^{\frac{D}{2}-1}} 
\,,
\ee
where $\bar \varphi(u,r)$ is given by a sum of terms proportional to
\be
\frac{\theta(u)}{(u(u+2r))^\alpha(u+r)^\beta}\,,
\ee
with $\alpha$, $\beta$ positive and $\alpha+\beta$ half odd. In particular, 
it is then clear that the limit of this field as $r\to\infty$ for any fixed $u$ does not display any term with the Coulombic behaviour $r^{3-D}$ and hence that there is no memory effect on $\mathscr I^+$ to that order, since 
\be
\int_0^{u(u+2r)}
\frac{d\kappa}{\sqrt{\kappa}(\kappa+r^2)^{\frac{D}{2}-1}} \sim \frac{1}{r^{D-3}}\int_0^{\frac{2u}{r}}\frac{dx}{\sqrt x(1+x)^{\frac{D}{2}-1}}\sim \frac{2\sqrt{2u}}{r^{D-\frac{5}{2}}}\,.
\ee 
Considering instead the limit of $\varphi$ as $t\to+\infty$ for fixed $r$, one sees that only the last term in \eqref{phioddD} survives and yields
\be
\frac{cq}{2}(-1)^{\frac{D-3}{2}}\int_0^{\infty}\frac{d\kappa}{\sqrt\kappa (\kappa+r^2)^{\frac{D}{2}-1}} = \frac{cq}{2}\frac{(-1)^{\frac{D-3}{2}}}{r^{D-3}}B\big(\tfrac{1}{2},\tfrac{D-3}{2}\big) = \frac{q}{(D-3)\Omega_{D-2}r^{D-3}}\,.
\ee
This means that the Coulombic energy due to the newly created particle is felt by the test charge only at $i^+$, namely after one has waited (for an infinite time) at a fixed distance $r$ that the perturbations due to the dispersion occurring in odd spacetime dimensions have died out. To some extent, this is to be regarded as a smeared-out memory effect, as opposed to memory effects occurring sharply at $\mathscr I^+$ near $u=0$ in even dimensions (see also \cite{SatishchandranWald}). 

The situation does not improve if one considers a particle that is created with a non-zero velocity $\mathbf v$. Indeed, boosting the exact solution \eqref{phioddD} by means of \eqref{standardboost}, one sees that $\varphi$ goes to zero for fixed $r$ as $t\to+\infty$. The reason is that, while one waits for the dispersion to die out, the source, moving at a constant velocity, has travelled infinitely far from the test charge. 

Shifting our attention to the case of null memory, we see that it is possible to provide the following formal solution to the recursion relations \eqref{recursionnullD}, which hold in any dimension. We consider $\varphi= \sum \varphi^{(k)}r^{-k}$, setting $\varphi^{(k)}=0$ for $k\ge D-3$, while, for $k \le D-4$,
\be\label{nullmemkodd}
\varphi^{(k)}(u, \mathbf n)=\delta^{(D-4-k)}(u)C_k(\mathbf n)\,,
\ee
with the functions $C_k(\mathbf n)$ determined recursively by
\be\begin{aligned}
	(D-2k-2) C_k(\mathbf n) &= [\Delta + (k-1)(k-D+2)]C_{k-1}(\mathbf n)\,,\\
	C_{D-4}(\mathbf n) &= -(\Delta-D+4)^{-1}(\mathbf n,{\mathbf {\hat x}})\,.
\end{aligned}\ee
Thus, although the field is highly singular at $u=0$, the resulting null memory effect will be formally identical to the one occurring in even dimensions.

\section{Exact solution of $\Box\epsilon=0$} \label{app:box}
The closed-form solution of the  wave equation \eqref{residual} with the boundary condition 
\be
\lim_{r\to\infty}\epsilon(u,r,\mathbf n) = \epsilon^{(0)}(u,\mathbf n)
\ee 
introduced in \eqref{boundaryfuncI} in any even dimension $D$ is given by 
\be\label{nostroprop}
\epsilon(x) = \frac{\Gamma(D-2)}{\pi^{\frac{D-2}{2}}\Gamma(\frac{D-2}{2})}\, \text{Re} \oint \frac{(-x^2)^{\frac{D-2}{2}}}{(-2x\cdot q+i\,\varepsilon)^{D-2}}\,\epsilon^{(0)}(\mathbf q)\, d\Omega(\mathbf q)\,,
\ee
where $q= (1,\mathbf q)$ and the limit $\varepsilon\to0^+$ is understood. The introduction of this small imaginary part is needed in order to avoid the singularities occurring in the angular integration for $|t|<|\mathbf x|$, namely outside the light-cone. Indeed, it is straightforward to verify that, for any value of $D$ even or odd,
\be
\Box\, \frac{(-x^2)^{\frac{D-2}{2}}}{(-2x\cdot q)^{D-2}} = 0\,,
\ee
while, aligning $\mathbf n$ along the $(D-1)$th direction, we have
\be\begin{aligned}
	&\mathrm{Re}
	\oint \frac{(-x^2)^{\frac{D-2}{2}}}{(-2x\cdot q+i\varepsilon)^{D-2}}\,\epsilon^{(0)}(\mathbf q)\, d\Omega(\mathbf q)
	\\
	&=\mathrm{Re}\int_{0}^\pi d\theta (\sin\theta)^{D-3}\oint d\Omega'(\mathbf q') \frac{2^{2-D}[u(u+2r)]^{\frac{D-2}{2}}}{[u+r(1-\cos\theta)+i\varepsilon]^{D-2}}\epsilon^{(0)}(\mathbf q' \sin\theta, \cos\theta)\,,
\end{aligned}
\ee
where $d\Omega'(\mathbf q')$ denotes the integral measure on the $(D-3)$-sphere. Letting $\tau=r(1-\cos\theta)/u$, for $u\neq0$, the previous expression becomes
\be\label{arccos_asymptD}
\text{Re}
\int_0^{2r/u}\frac{u \,d\tau}{r}\,\frac{(1+\frac{2r}{u})^{\frac{D-2}{2}}(
	\tfrac{2u\tau}{r}-\tfrac{u^2\tau^2}{r^2})^{\frac{D-4}{2}}}{2^{D-2}(1+\tau+i\varepsilon)^{D-2}}
\oint d\Omega'(\mathbf q')\, 
\epsilon^{(0)}\left(
\mathbf q' \sqrt{
	\tfrac{2u\tau}{r}-\tfrac{u^2\tau^2}{r^2}}
,
1-\tfrac{u \tau}{r}
\right)\,,
\ee
which, as $r\to\infty$, tends to 
\be
\frac{1}{2}\,
\mathrm{Re} \int_0^{u\cdot\infty} \frac{\tau^{\frac{D-4}{2}}d\tau}{(1+\tau + i\varepsilon)^{D-2}}
\oint d\Omega'(\mathbf q') \epsilon^{(0)}(\mathbf n)=\frac{\pi^{\frac{D-2}{2}}\Gamma(\frac{D-2}{2})}{\Gamma({D-2})}\,\epsilon^{(0)}(\mathbf n)\,.
\ee
The solution \eqref{nostroprop} which, to the best of our knowledge, was not previously exhibited in closed form in the literature, is thus compatible with the asymptotic expansion \eqref{expepsilonlog} and generalises the expression given in \cite{Hirai-Sugishita}.

In addition, being non-perturbative, it allows to explicitly verify the antipodal matching condition:
\be
\lim_{r\to\infty}\epsilon(u=v-2r, r,\mathbf n)=\epsilon^{(0)}(-\mathbf n)
\ee
for any fixed advanced time $v$.

\section{Soft photon theorem in even $D$}\label{app:Weinberg}

We would like to show that the surface charges defined in Section \ref{ssec:Finitecharges} enter Weinberg's soft theorem \cite{Weinberg64, Weinberg65}. More precisely, we will see how the Weinberg theorem implies the validity of the Ward identities associated to such charges in $D\geq4$ \cite{MitraHePhoton, MitraHeGauge}. 

The surface charge associated to \eqref{expepsilonlog}, evaluated at $\mathscr I^+_-$ reads
\be\label{Strominger1Charge}
\mathcal Q_\epsilon = \int_{\mathscr I^+_-}
\left(
\partial_u A^{(D-2)}_r+(D-3)A_u^{(D-3)}
\right)
\epsilon^{(0)}d\Omega_{D-2}\,,
\ee
where we have taken into account the absence of radiation terms for $u\to-\infty$. Recasting \eqref{Strominger1Charge} as an integral over the whole of $\mathscr I^+$, and assuming that no contribution arises at $\mathscr I^+_+$, which is the case in particular if there are no stable massive charge, we find 
\be\label{Strominger-charge}
\mathcal Q_\epsilon= -\frac{1}{r^{D-4}}\int_{\mathscr I^+} \partial_u^2 A_r^{(D-2)} \epsilon^{(0)} du\, d\Omega_{D-2}\,,
\ee 
where we have used the fact that, after recursive gauge fixing, $A_u^{(D-3)}$ is independent of $u$ on shell.
We would like to express \eqref{Strominger-charge} in terms of the leading radiation field, which, as we shall see below, indeed contains the creation and annihilation operator of asymptotic photons.
To this end, we first combine \eqref{eqr_r} and \eqref{Lorenz_r} and obtain
\be
\cD\cdot A^{(k-1)} = \frac{\Delta-(D-2-k)(D-3-k)}{D-2-2k}A_r^{(k)}+(D-3-k)A_u^{(k)}\,;
\ee
employing \eqref{equ_r} as well,
\be
\partial_u A_r^{(k+1)} = \mathscr D_k A_r^{(k)} - A_u^{(k)}\,,
\ee
where $\mathscr D_k$ is given in \eqref{operatoresfera}.

Employing this relation recursively, we find
\be\label{recursive-u}
\partial_u^{D/2} A_r^{(D-2)} = \prod_{l=D/2}^{D-3} \mathscr D_l\,\, \partial_u \cD\cdot A^{(\frac{D-4}{2})}\,,
\ee
where we have used \eqref{eqr_r} to deduce $\partial_u A_r^{(\frac{D}{2})} = \cD\cdot A^{(\frac{D-4}{2})}$. In the above writing, we adopt the convention that for $D=4$ the product $\prod_l \mathscr D_l$ (which in this case has a formally ill-defined range) reduces to the identity. We can then use \eqref{recursive-u} to recast \eqref{Strominger-charge} as
\be
\mathcal Q_\epsilon = -\frac{1}{r^{D-4}}\int_{-\infty}^{+\infty}\left(\int_{-\infty}^u du\right)^{D/2-2} \partial_u \cD\cdot A^{(\frac{D-4}{2})} \prod_{l=D/2}^{D-3}\mathscr D_l \epsilon^{(0)}du\,d\Omega_{D-2}\,. 
\ee
On the other hand, the asymptotic expansion of the free electromagnetic field operator, expressed in terms of creation and annihilation operators, yields, to leading order,
\be
\mathcal A_i(u,r,x^k) = \frac{i^{1-D/2}}{8\pi^{2}r^{(D-4)/2}} \int_0^{+\infty} \left(\frac{\omega}{2\pi}\right)^{(D-4)/2}e^{-i\omega u} \epsilon^\sigma_i (\hat x) a_\sigma(\omega\hat x)\,d\omega+h.c.\,,
\ee
where
$\hat x^\mu = x^\mu/r$, while $\epsilon^\sigma$ are polarisation tensors for the $D-2$ propagating helicities. This formula provides thus an explicit expression for $A_i^{(\frac{D-4}{2})}$ and hence allows us to make explicit the relation between the charge $\mathcal Q_\epsilon$ and the soft photon creation and annihilation operators as follows (we employ the prescription $\int_{-\infty}^{+\infty}du \int_{0}^{+\infty}d\omega\, e^{i\omega u} f(\omega)=\lim_{\omega\to 0^+}f(\omega)/2$) 
\be
\mth Q_\epsilon = \frac{1}{8(2\pi)^{(D-2)/2} r^{D-4}} \lim_{\omega\to0^+} \int_{S^{D-2}}
\cD^i[\epsilon_i^\sigma(\hat x) \omega a_\sigma(\omega\hat x) + h.c.] 
\prod_{l=D/2}^{D-3} \mathscr D_l\,\, \epsilon^{(0)}(\hat x) d\Omega_{D-2}(\hat x)\,.
\ee
Assuming that the charge $\mathcal Q_\epsilon$, together with its counterpart at $\mathscr I^-$, generates the residual symmetry $\delta \psi(u,r,\hat x) = i\epsilon^{(0)}(\hat x)+\mathcal O(r^{-1})$ in a canonical way, and employing suitable antipodal matching and crossing symmetry conditions, we have the Ward identity
\beal\label{Ward-identity-per-Weinberg}
&\frac{1}{2(2\pi)^{(D-2)/2}} \int_{S^{D-2}}  \epsilon_i^\sigma(\hat x) \lim_{\omega\to0^+} \cD^i\langle\text{out}|\omega a_\sigma(\omega\hat x) \mathcal S |\text{in}\rangle  \prod_{l=D/2}^{D-3} \mathscr D_l\,\, \epsilon^{(0)}(\hat x)\, d\Omega_{D-2}(\hat x) \\ 
&= 
\sum_{n} e_n \epsilon^{(0)}(\hat x_n) \langle\text{out}| \mathcal S |\text{in}\rangle,
\eeal
where the sum on the right-hand side extends to all charged external particles in the amplitude and $e_n$ is the electric charge of the $n$th particle (taking into account with a suitable sign whether the particle is outgoing, resp. incoming).
Notably, the left-hand side contains exactly the combination $\mathcal P[\,\cdot\,] = \lim_{\omega\to 0^+}[\omega\,\cdot\,]$ that selects the pole in the amplitude with the soft insertion.

On the other hand, the Weinberg theorem for an amplitude involving external massless particles with momenta $p_n=E_n(1,\hat x_n)$ and a soft photon emitted with helicity $\sigma$ pointing along the $\hat n$ direction on the celestial sphere reads
\be\label{Weinberg-massless}
\lim_{\omega\to0^+} \langle \text{out}| \omega a_{\sigma}(\omega \hat n) \mathcal S |\text{in}\rangle = \sum_n \frac{e_n \epsilon^\sigma(\hat n)^\ast\cdot p_n}{p_n\cdot(1,\hat n)}
\langle \text{out}| \mathcal S |\text{in} \rangle\,.
\ee
Multiplying this relation by $\epsilon^\sigma_i(\hat n)$ and summing over $\sigma$, we see that this is equivalent to
\be\label{identity_i}
\epsilon^{\sigma}_i(\hat n) 
\lim_{\omega\to0^+}\langle \text{out}|\omega a_\sigma(\omega \hat n) S |\text{in}\rangle
=\sum_{n}e_n \cD_i \alpha(\hat x_n, \hat n) \langle \text{out}| S |\text{in}\rangle\,,
\ee
where we have used the completeness relation for polarisation vectors and defined a function
\be
\alpha(\hat x, \hat n) = \log(1-\hat x \cdot \hat n)\,.
\ee
This function $\alpha$ satisfies
the following identity (see \cite{StromingerQEDevenD})
\be\label{delta-identity}
\frac{1}{2(2\pi)^{(D-2)/2}}\prod_{l=D/2}^{D-3}\mathscr D_l\,\, \Delta \alpha(\hat x, \hat n) = \delta(\hat x, \hat n)\,,
\ee
where $\hat x$ is here treated as a constant vector on the unit sphere and
$\delta(\hat x, \hat n)$ is the invariant delta function on the $(D-2)$-sphere. Now, acting with the differential operator 
\be
\frac{1}{2(2\pi)^{(D-2)/2}}\prod_{l=D/2}^{D-3}\mathscr D_l\,\cdot \cD^i
\ee
on equation \eqref{identity_i}, multiplying by an arbitrary $\epsilon^{(D-4)}(\hat x)$ and integrating over the unit sphere allows one to retrieve the Ward identity \eqref{Ward-identity-per-Weinberg}, thanks to the relation \eqref{delta-identity}. This proves that the Weinberg factorisation implies the existence of the asymptotic symmetry Ward identities. 

Remarkably, the charge \eqref{ChargeStrom2} associated to the symmetry \eqref{res_even}, which is responsible for the memory effect, formally differs from \eqref{Strominger1Charge} only by a factor of $1/r^{D-4}$ (other than by the substitution $\epsilon^{(0)}\leftrightarrow\epsilon^{(D-4)}$), which makes it vanish on $\mathscr I^+$. However, the corresponding symmetry transformation of the matter fields would be $\delta \psi(u,r,\hat x) = i\epsilon^{(D-4)}(\hat x)/r^{D-4}+\mathcal O(r^{3-D})$, and hence would give rise to Ward identities completely equivalent to \eqref{Ward-identity-per-Weinberg}, with the factors of $1/r^{D-4}$ cancelling each other on the two sides. This indicates that the large gauge symmetry \eqref{expansioneq-with-logs} and the residual symmetry \eqref{res_even}, acting at Coulombic order, can both be seen as consequences of Weinberg's soft theorem. This is also reflected in the observation that the Fourier transform of the soft factor occurring in Weinberg's theorem is strictly related to the memory formulae \cite{Mao-Ouyang}.

%%%%%%%%%%%%%%%%%%%%%%%%%%%%%%%%%%%%%%%%%%%%%%%%%%


\begin{thebibliography}{100}

\bibitem{ZP}
Y. B. Zeldovich and A. G. Polnarev, 
``{\it Radiation of gravitational waves by a cluster of superdense stars},''
Sov.\ Astron.\ {\bf 18} (1974) 17.

\bibitem{Christodoulou:1991cr}
  D.~Christodoulou,
 ``{\it Nonlinear nature of gravitation and gravitational wave experiments},''
  Phys.\ Rev.\ Lett.\  {\bf 67} (1991) 1486.
  %doi:10.1103/PhysRevLett.67.1486
  %%CITATION = doi:10.1103/PhysRevLett.67.1486;%%

\bibitem{BG2013}
  L.~Bieri and D.~Garfinkle,
  ``{\it An electromagnetic analogue of gravitational wave memory},''
  Class.\ Quant.\ Grav.\  {\bf 30} (2013) 195009
 % doi:10.1088/0264-9381/30/19/195009
  [\href{https://arxiv.org/abs/1307.5098}{[arXiv:1307.5098 [gr-qc]}].
  %%CITATION = doi:10.1088/0264-9381/30/19/195009;%%
 
\bibitem{Tolish:2014bka}
  A.~Tolish and R.~M.~Wald,
  ``\textit{Retarded Fields of Null Particles and the Memory Effect},''
  Phys. Rev. D {\bf 89} (2014)  064008
  %doi:10.1103/PhysRevD.89.064008
  [\href{https://arxiv.org/abs/1401.5831}{arXiv:1401.5831 [gr-qc]}].
  %%CITATION = doi:10.1103/PhysRevD.89.064008;%%
  
\bibitem{Pasterski:2015zua}
  S.~Pasterski,
 ``{\it Asymptotic Symmetries and Electromagnetic Memory},''
  JHEP {\bf 1709} (2017) 154
%  doi:10.1007/JHEP09(2017)154
  [\href{https://arxiv.org/abs/1505.00716}{[arXiv:1505.00716 [hep-th]}].
  %%CITATION = doi:10.1007/JHEP09(2017)154;%% 

\bibitem{MaoetAl}
P.~Mao, H.~Ouyang, J.~B.~Wu and X.~Wu,
``\textit{New electromagnetic memories and soft photon theorems},''
Phys.\ Rev.\ D {\bf 95} (2017) 125011
[\href{https://arxiv.org/abs/1703.06588}{arXiv:1703.06588 [hep-th]}].
%%CITATION = doi:10.1103/PhysRevD.95.125011;%%

\bibitem{Sarkkinentesi}
M.~Sarkkinen,
``\textit{Memory effect in electromagnetic radiation},'' \href{https://helda.helsinki.fi/handle/10138/231490}{HELDA}.
%%CITATION = INSPIRE-1676650;%%

\bibitem{Sarkkinenpaper}
N.~Jokela, K.~Kajantie and M.~Sarkkinen,
``\textit{Memory effect in Yang-Mills theory},''
Phys.\ Rev.\ D {\bf 99} (2019)  116003
[\href{https://arxiv.org/abs/1903.10231}{arXiv:1903.10231 [hep-th]}].
%%CITATION = doi:10.1103/PhysRevD.99.116003;%%
    
\bibitem{Garfinkle:2017fre}
  D.~Garfinkle, S.~Hollands, A.~Ishibashi, A.~Tolish and R.~M.~Wald,
  ``{\it The Memory Effect for Particle Scattering in Even Spacetime Dimensions},''
  Class.\ Quant.\ Grav.\  {\bf 34} (2017) 145015
%  doi:10.1088/1361-6382/aa777b
  \href{https://arxiv.org/abs/1702.00095}{[arXiv:1702.00095 [gr-qc]}].
  %%CITATION = doi:10.1088/1361-6382/aa777b;%%

\bibitem{Mao-Ouyang}
  P.~Mao and H.~Ouyang,
  ``{\it Note on soft theorems and memories in even dimensions},''
  Phys.\ Lett.\ B {\bf 774} (2017) 715
%  doi:10.1016/j.physletb.2017.08.064
  [\href{https://arxiv.org/abs/1707.07118}{arXiv:1707.07118 [hep-th]}].
  %%CITATION = doi:10.1016/j.physletb.2017.08.064;%%

\bibitem{HamadaSeoShiu}
Y.~Hamada, M.~S.~Seo and G.~Shiu,
``\textit{Electromagnetic Duality and the Electric Memory Effect},''
JHEP {\bf 1802} (2018) 46
[\href{https://arxiv.org/abs/1711.09968}{arXiv:1711.09968 [hep-th]}].
%%CITATION = doi:10.1007/JHEP02(2018)046;%%
%10 citations counted in INSPIRE as of 13 Jun 2019 
  
  \bibitem{WaldOddD}
  G.~Satishchandran and R.~M.~Wald,
 ``\textit{Memory effect for particle scattering in odd spacetime dimensions},''
  Phys.\ Rev.\ D {\bf 97} (2018) 24036
  [\href{https://arxiv.org/abs/1712.00873}{arXiv:1712.00873 [gr-qc]}].
  %%CITATION = doi:10.1103/PhysRevD.97.024036;%%
  %9 citations counted in INSPIRE as of 29 Apr 2019

\bibitem{HamadaSugishita}
Y.~Hamada and S.~Sugishita,
``{\it Notes on the gravitational, electromagnetic and axion memory effects},''
JHEP {\bf 1807} (2018) 17
%doi:10.1007/JHEP07(2018)017
[\href{https://arxiv.org/abs/1803.00738}{arXiv:1803.00738 [hep-th]}].
%%CITATION = doi:10.1007/JHEP07(2018)017;%%
  
 \bibitem{ShahinString}
 H.~Afshar, E.~Esmaeili and M.~M.~Sheikh-Jabbari,
 ``\textit{String Memory Effect},''
 JHEP {\bf 1902} (2019) 53
 [\href{https://arxiv.org/abs/1811.07368}{arXiv:1811.07368 [hep-th]}].
 %%CITATION = doi:10.1007/JHEP02(2019)053;%%
 %3 citations counted in INSPIRE as of 13 Jun 2019
  
%\cite{Satishchandran:2019pyc}
\bibitem{SatishchandranWald}
G.~Satishchandran and R.~M.~Wald,
``{\it Asymptotic behavior of massless fields and the memory effect},''
Phys.\ Rev.\ D {\bf 99} (2019) 84007
%doi:10.1103/PhysRevD.99.084007
[\href{https://arxiv.org/abs/1901.05942}{arXiv:1901.05942 [gr-qc]}].
%%CITATION = doi:10.1103/PhysRevD.99.084007;%%
%3 citations counted in INSPIRE as of 29 Apr 2019

\bibitem{Strominger2}
M.~Pate, A.~M.~Raclariu and A.~Strominger,
``\textit{Gravitational Memory in Higher Dimensions},''
JHEP {\bf 1806} (2018) 138
[\href{https://arxiv.org/abs/1712.01204}{arXiv:1712.01204 [hep-th]}].
%%CITATION = doi:10.1007/JHEP06(2018)138;%%

\bibitem{Susskind:2015hpa}
L.~Susskind,
``{\it Electromagnetic Memory},''
\href{https://arxiv.org/abs/1507.02584}{arXiv:1507.02584 [hep-th]}.
  %%CITATION = ARXIV:1507.02584;%%
  %34 citations counted in INSPIRE as of 13 Jun 2019 

\bibitem{StromingerC}
  M.~Pate, A.~M.~Raclariu and A.~Strominger,
  ``{\it Color Memory: A Yang-Mills Analog of Gravitational Wave Memory},''
  Phys.\ Rev.\ Lett.\  {\bf 119} (2017) 261602
%  doi:10.1103/PhysRevLett.119.261602
  [\href{https://arxiv.org/abs/1707.08016}{arXiv:1707.08016 [hep-th]}].
  %%CITATION = doi:10.1103/PhysRevLett.119.261602;%%
  
\bibitem{Strominger_YM} 
A.~Strominger,
  ``{\it Asymptotic Symmetries of Yang-Mills Theory},''
  JHEP {\bf 1407} (2014) 151
  %doi:10.1007/JHEP07(2014)151
  [\href{http://arxiv.org/abs/arXiv:1308.0589}{arXiv:1308.0589 [hep-th]}].
  %%CITATION = doi:10.1007/JHEP07(2014)151;%%
  
\bibitem{BarnichYM}
G.~Barnich and P.~H.~Lambert,
``\textit{Einstein-Yang-Mills theory: Asymptotic symmetries},''
Phys.\ Rev.\ D {\bf 88} (2013) 103006
[\href{https://arxiv.org/abs/1310.2698}{arXiv:1310.2698 [hep-th]}].
%%CITATION = doi:10.1103/PhysRevD.88.103006;%%

\bibitem{Strominger_QED} 
T.~He, P.~Mitra, A.~P.~Porfyriadis and A.~Strominger,
  ``{\it New Symmetries of Massless QED},''
  JHEP {\bf 1410} (2014) 112
  %doi:10.1007/JHEP10(2014)112
  [\href{http://arxiv.org/abs/arXiv:1407.3789}{arXiv:1407.3789 [hep-th]}].
  %%CITATION = doi:10.1007/JHEP10(2014)112;%%
  
\bibitem{soft_QED_Strominger} 
V.~Lysov, S.~Pasterski and A.~Strominger,
``{\it Low's Subleading Soft Theorem as a Symmetry of QED},''
Phys.\ Rev.\ Lett.\  {\bf 113} (2014) 111601
%doi:10.1103/PhysRevLett.113.111601
[\href{http://arxiv.org/abs/arXiv:1407.3814}{arXiv:1407.3814 [hep-th]].}
%%CITATION = doi:10.1103/PhysRevLett.113.111601;%%

\bibitem{StromingerQEDevenD}
D.~Kapec, V.~Lysov and A.~Strominger,
``\textit{Asymptotic Symmetries of Massless QED in Even Dimensions},''
Adv.\ Theor.\ Math.\ Phys.\  {\bf 21} (2017) 1747
[\href{https://arxiv.org/abs/1412.2763}{arXiv:1412.2763 [hep-th]}].
%%CITATION = doi:10.4310/ATMP.2017.v21.n7.a6;%%
%51 citations counted in INSPIRE as of 29 Apr 2019

\bibitem{Campiglia} 
M.~Campiglia and A.~Laddha,
  ``{\it Asymptotic symmetries of QED and Weinberg's soft photon theorem},''
  JHEP {\bf 1507} (2015) 115
  %doi:10.1007/JHEP07(2015)115
  [\href{http://arxiv.org/abs/arXiv:1505.05346}{arXiv:1505.05346 [hep-th]}].
  %%CITATION = doi:10.1007/JHEP07(2015)115;%%
  
 \bibitem{StromingerMagnetic}
 A.~Strominger,
 ``\textit{Magnetic Corrections to the Soft Photon Theorem},''
 Phys.\ Rev.\ Lett.\  {\bf 116} (2016)  031602
 [\href{https://arxiv.org/abs/1509.00543}{arXiv:1509.00543 [hep-th]}].
 %%CITATION = doi:10.1103/PhysRevLett.116.031602;%%

\bibitem{Avery_Schwab} 
S.~G.~Avery and B.~U.~W.~Schwab,
  ``{\it Noether's second theorem and Ward identities for gauge symmetries},''
  JHEP {\bf 1602} (2016) 31
  %doi:10.1007/JHEP02(2016)031
  [\href{http://arxiv.org/abs/arXiv:1510.07038}{arXiv:1510.07038 [hep-th]}].
  %%CITATION = doi:10.1007/JHEP02(2016)031;%%
  
\bibitem{ACD2}
  A.~Campoleoni, D.~Francia and C.~Heissenberg,
  ``\textit{Asymptotic Charges at Null Infinity in Any Dimension},''
  Universe {\bf 4} (2018) 47
  [\href{https://arxiv.org/abs/1712.09591}{arXiv:1712.09591 [hep-th]}].
  %%CITATION = doi:10.3390/universe4030047;%%
  %11 citations counted in INSPIRE as of 13 Jun 2019
    
\bibitem{Henneaux:2018gfi}
  M.~Henneaux and C.~Troessaert,
  ``{\it Asymptotic symmetries of electromagnetism at spatial infinity},''
  JHEP {\bf 1805} (2018) 137
%  doi:10.1007/JHEP05(2018)137
  [\href{https://arxiv.org/abs/1803.10194}{arXiv:1803.10194 [hep-th]}].
  %%CITATION = doi:10.1007/JHEP05(2018)137;%%
  
\bibitem{Hirai-Sugishita}
H.~Hirai and S.~Sugishita,
``\textit{Conservation Laws from Asymptotic Symmetry and Subleading Charges in QED},''
JHEP {\bf 1807} (2018) 122
%doi:10.1007/JHEP07(2018)122
[\href{https://arxiv.org/abs/1805.05651}{arXiv:1805.05651 [hep-th]}].
%%CITATION = doi:10.1007/JHEP07(2018)122;%%
%6 citations counted in INSPIRE as of 29 Apr 2019

\bibitem{ShahinMagnetic}
V.~Hosseinzadeh, A.~Seraj and M.~M.~Sheikh-Jabbari,
``\textit{Soft Charges and Electric-Magnetic Duality},''
JHEP {\bf 1808} (2018) 102
[\href{https://arxiv.org/abs/1806.01901}{arXiv:1806.01901 [hep-th]}].
%%CITATION = doi:10.1007/JHEP08(2018)102;%%

\bibitem{Campoleoni:2018uib}
  A.~Campoleoni, D.~Francia and C.~Heissenberg,
  ``\textit{Asymptotic symmetries and charges at null infinity: from low to high spins},''
  EPJ Web Conf.\  {\bf 191} (2018) 06011
  %doi:10.1051/epjconf/201819106011
  [\href{https://arxiv.org/abs/1808.01542}{arXiv:1808.01542 [hep-th]}].
  
 \bibitem{Tristan}
 R.~Gonzo, T.~McLoughlin, D.~Medrano and A.~Spiering,
 ``\textit{Asymptotic Charges and Coherent States in QCD},''
 \href{https://arxiv.org/abs/1906.11763}{arXiv:1906.11763 [hep-th]}.
 %%CITATION = ARXIV:1906.11763;%%

\bibitem{MitraHePhoton}
T.~He and P.~Mitra,
``\textit{Asymptotic Symmetries and Weinberg's Soft Photon Theorem in Mink$_{d+2}$},''
\href{https://arxiv.org/abs/1903.02608}{arXiv:1903.02608 [hep-th]}.
%%CITATION = ARXIV:1903.02608;%%
%3 citations counted in INSPIRE as of 13 Jun 2019

\bibitem{MitraHeGauge}
T.~He and P.~Mitra,
``\textit{Asymptotic Symmetries in $(d+2)$-Dimensional Gauge Theories},''
\href{https://arxiv.org/abs/1903.03607}{arXiv:1903.03607 [hep-th]}.
%%CITATION = ARXIV:1903.03607;%%

\bibitem{MarcCedric-anyD}
  M.~Henneaux and C.~Troessaert,
  ``{\it Asymptotic structure of electromagnetism in higher spacetime dimensions},''
  Phys.\ Rev.\ D {\bf 99} (2019) 125006
%  doi:10.1103/PhysRevD.99.125006
  [\href{https://arxiv.org/abs/1903.04437}{arXiv:1903.04437 [hep-th]}].
  %%CITATION = doi:10.1103/PhysRevD.99.125006;%%

\bibitem{He:2019esa}
  T.~He and P.~Mitra,
  ``{\it New Magnetic Symmetries in $(d+2)$-Dimensional QED},''
  \href{https://arxiv.org/abs/1907.02808}{arXiv:1907.02808 [hep-th]}.
  %%CITATION = ARXIV:1907.02808;%%
  
\bibitem{Lectures}
 A.~Strominger,
 ``\textit{Lectures on the Infrared Structure of Gravity and Gauge Theory},'' Princeton University Press, Princeton, 2018
 [\href{https://arxiv.org/abs/1703.05448}{arXiv:1703.05448 [hep-th]}].
 %%CITATION = ARXIV:1703.05448;%%
 %215 citations counted in INSPIRE as of 11 Jul 2019
  
%\cite{Wong:1970fu}
\bibitem{WongEquations}
  S.~K.~Wong,
  ``{\it Field and particle equations for the classical Yang-Mills field and particles with isotopic spin},''
  Nuovo Cim.\ A {\bf 65} (1970) 689.
  %%CITATION = doi:10.1007/BF02892134;%%
  %418 citations counted in INSPIRE as of 13 Jun 2019
  
\bibitem{LibroRusso}
	B.~Kosyakov,
  ``{\it Introduction to the Classical Theory of Particles and Fields},'' Springer-Verlag, Berlin-Heidelberg, 2007.
   
 \bibitem{alternative-bnd}
 D.~Kapec, V.~Lysov, S.~Pasterski and A.~Strominger,
 ``\textit{Higher-Dimensional Supertranslations and Weinberg's Soft Graviton Theorem},''
 Ann. Math. Sci. Appl. \textbf{2} (2017)
 69
 %doi:10.4310/AMSA.2017.v2.n1.a2
 [\href{http://arxiv.org/abs/arXiv:1502.07644}{arXiv:1502.07644 [gr-qc]}].
 %%CITATION = doi:10.4310/AMSA.2017.v2.n1.a2;%%
 %66 citations counted in INSPIRE as of 08 Jul 2019
 
\bibitem{Campiglia-Soni}
M.~Campiglia, L.~Freidel, F.~Hopfmueller and R.~M.~Soni,
``\textit{Scalar Asymptotic Charges and Dual Large Gauge Transformations},''
JHEP {\bf 1904} (2019) 3
[\href{https://arxiv.org/abs/1810.04213}{arXiv:1810.04213 [hep-th]}].
%%CITATION = doi:10.1007/JHEP04(2019)003;%%
%7 citations counted in INSPIRE as of 09 Jul 2019
 
\bibitem{CD2form}
D.~Francia and C.~Heissenberg,
``\textit{Two-Form Asymptotic Symmetries and Scalar Soft Theorems},''
Phys.\ Rev.\ D {\bf 98} (2018) 105003
[\href{https://arxiv.org/abs/1810.05634}{arXiv:1810.05634 [hep-th]}].
%%CITATION = doi:10.1103/PhysRevD.98.105003;%%
%4 citations counted in INSPIRE as of 13 Jun 2019

\bibitem{Riello}
A.~Riello,
``\textit{Soft charges from the geometry of field space},''
\href{https://arxiv.org/abs/1904.07410}{arXiv:1904.07410 [hep-th]}.
%%CITATION = ARXIV:1904.07410;%%
%3 citations counted in INSPIRE as of 22 Jun 2019
  
\bibitem{Aggarwal}
A.~Aggarwal,
``\textit{Supertranslations in Higher Dimensions Revisited},''
Phys. Rev. D {\bf 99} (2019) 026015
%doi:10.1103/PhysRevD.99.026015
[\href{https://arxiv.org/abs/1811.00093}{arXiv:1811.00093 [hep-th]}].
%%CITATION = doi:10.1103/PhysRevD.99.026015;%%

\bibitem{FreidelHopfmuellerRiello}
L.~Freidel, F.~Hopfm\"uller and A.~Riello,
``\textit{Holographic Renormalization in Flat Space: Symplectic Potential and Charges of Electromagnetism},''
\href{https://arxiv.org/abs/1904.04384}{arXiv:1904.04384 [hep-th]}.
%%CITATION = ARXIV:1904.04384;%%
%1 citations counted in INSPIRE as of 29 Apr 2019

\bibitem{Barnich-Brandt}
G.~Barnich and F.~Brandt,
``{\it Covariant theory of asymptotic symmetries, conservation laws and central charges},''
Nucl.\ Phys.\ B {\bf 633} (2002) 3
%doi:10.1016/S0550-3213(02)00251-1
[\href{https://arxiv.org/abs/hep-th/0111246}{hep-th/0111246}].
%%CITATION = doi:10.1016/S0550-3213(02)00251-1;%%
%377 citations counted in INSPIRE as of 29 Apr 2019

\bibitem{Friedlander}
F.~G.~Friedlander, ``\textit{The Wave Equation on a Curved Space-Time},''
	Cambridge University Press,
	Cambridge, 1975.
	
\bibitem{Weinberg64}
S.~Weinberg,
``{\it Photons and Gravitons in s Matrix Theory: Derivation of Charge Conservation and Equality of Gravitational and Inertial Mass},''
Phys. Rev. B {\bf 135} (1964) 1049.
%doi:10.1103/PhysRev.135.B1049
%%CITATION = doi:10.1103/PhysRev.135.B1049;%%
%427 citations counted in INSPIRE as of 08 Jul 2019

\bibitem{Weinberg65}
S.~Weinberg,
``{\it Infrared photons and gravitons},''
Phys. Rev. B {\bf 140} (1965) 516.
%doi:10.1103/PhysRev.140.B516
%%CITATION = doi:10.1103/PhysRev.140.B516;%%
%621 citations counted in INSPIRE as of 08 Jul 2019

\end{thebibliography}
\end{document}